\documentclass{IEEEtaes}

\usepackage{color,array,amsthm}
\usepackage{graphicx}

\usepackage{cite}
\usepackage{makecell}
\usepackage{amsthm}
\usepackage{amsmath}
\usepackage[euler]{textgreek}
\usepackage[linesnumbered,ruled,vlined]{algorithm2e}
\usepackage{array}

\usepackage{xcolor}
\SetKwInput{KwInput}{Input}                % Set the Input
\SetKwInput{KwOutput}{Output}              % set the Output

\SetCommentSty{mycommfont}
\ifCLASSOPTIONcompsoc
  \usepackage[caption=false,font=normalsize,labelfont=sf,textfont=sf]{subfig}
\else
  \usepackage[caption=false,font=footnotesize]{subfig}
\fi

\usepackage{graphics}

\begin{document}

\title{Bayesian Linear Regression with Cauchy Prior and Its Application in Sparse MIMO Radar}

\author{Jun Li}
\member{Member, IEEE}
\affil{NXP Semiconductors, San Jose, CA, USA} 

\author{Ryan Wu}
\member{Member, IEEE}
\affil{NXP Semiconductors, San Jose, CA, USA} 

\author{I-Tai Lu}
\member{Senior Member, IEEE}
\affil{New York University, NY, USA}

\author{Dongyin Ren}
\member{Member, IEEE}
\affil{NXP Semiconductors, San Jose, CA, USA}

%% \author{FOURTH D. AUTHOR}
%% \affil{University of Colorado, Colorado, USA}

\receiveddate{Manuscript received XXXXX 00, 0000; revised XXXXX 00, 0000; accepted XXXXX 00, 0000.\\
This work was supported by NXP Semiconductors, CA, USA\\}
\corresp{{\itshape (Corresponding author: Jun Li)}}
%This paragraph of the first footnote will contain the date on which you submitted your paper for review, which is populated by IEEE. It is IEEE style to display support information, including sponsor and financial support acknowledgment, here and not in an acknowledgment section at the end of the article. For example, ``This work was supported in part by the U.S. Department of Commerce under Grant BS123456.'' }
%% \accepteddate{XXXXX XX XXXX}
%% \publisheddate{XXXXX XX XXXX}

\authoraddress{Author’s addresses: Jun Li, Ryan Wu, and Dongyin Ren are with NXP Semiconductors, San Jose, CA, USA. Jun Li was formerly with NYU WIRELESS and the Department of Electrical and Computer Engineering, Tandon School of Engineering, New York University, NY, USA (e-mail: \href{mailto:jl7333@nyu.edu}{jl7333@nyu.edu}; \href{mailto:ryan.wu@nxp.com}{ryan.wu@nxp.com}; \href{mailto:dongyin.ren@nxp.com}{dongyin.ren@nxp.com}). I-Tai Lu is with the Department of Electrical and Computer Engineering, Tandon School of Engineering, New York University, NY, USA (e-mail: \href{mailto:itl211@nyu.edu}{itl211@nyu.edu}).}

%\editor{Mentions of supplemental materials and animal/human rights statements can be included here.}
%\supplementary{Color versions of one or more of the figures in this article are available online at \href{http://ieeexplore.ieee.org}{http://ieeexplore.ieee.org}.}

\markboth{Li ET AL.}{BAYESIAN LINEAR REGRESSION WITH CAUCHY PRIOR}
\maketitle

\begin{abstract}In this paper, a sparse signal recovery algorithm using Bayesian linear regression with Cauchy prior (BLRC) is proposed. Utilizing an approximate expectation maximization (AEM) scheme, a systematic hyper-parameter updating strategy is developed to make BLRC practical in highly dynamic scenarios. Remarkably, with a more compact latent space, BLRC not only possesses essential features of the well-known sparse Bayesian learning (SBL) and iterative reweighted $l_2$ (IR-$l_2$) algorithms but also outperforms them. Using sparse array (SPA) and coprime array (CPA), numerical analyses are first performed to show the superior performance of BLRC under various noise levels, array sizes, and sparsity levels. Applications of BLRC to sparse multiple-input and multiple-output (MIMO) radar array signal processing are then carried out to show that the proposed BLRC can efficiently produce high-resolution images of the targets.
\end{abstract}


\section{INTRODUCTION}
With the development of modern advanced driver-assistance systems (ADAS) and autonomous driving (AD) applications, accurate perception and interpretation of the surrounding environment are highly desirable for automotive radar systems. To satisfy the stringent perception and interpretation requirement, high-resolution automotive radar is being developed to provide point cloud of the surrounding environment in four-dimensions (4D), i.e., range, Doppler, and azimuth and elevation angles \cite{engels2021automotive}.

\begin{figure*}[!t]
\centering
\includegraphics[width=6.8in]{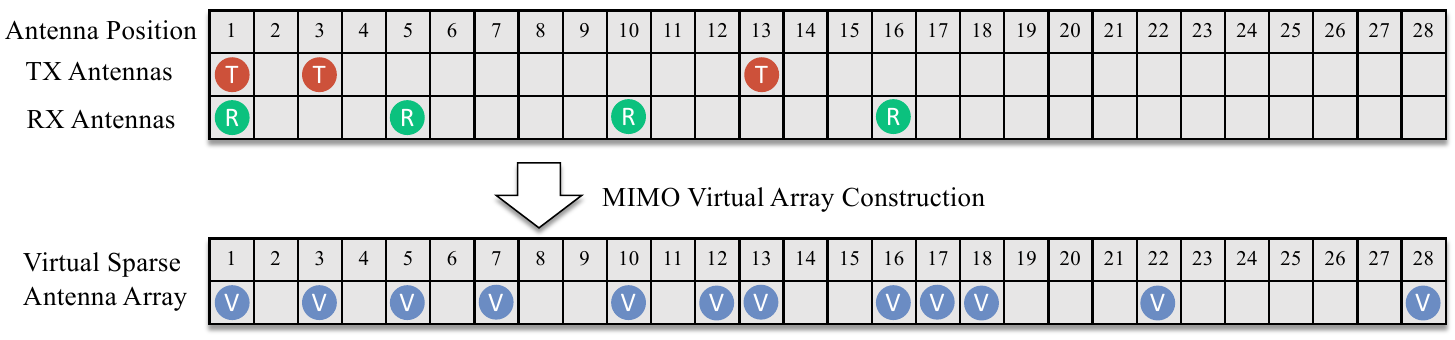}
\caption{Forming a sparse array of 12 virtual elements spanning an aperture of the size of 28 elements using a 3-transmitter/4-receiver MIMO radar system.}
\label{fig:virtual_array}
\end{figure*}

In modern automotive radar system, the Size, Weight, Power, and Cost (SWaP-C) requirement needs to be satisfied. For instance, the size of automotive radar sensor array has to be small enough to be placed behind the vehicle bumper. Thus, it is obvious that the resolution of automotive radar cannot increase easily due to the size constraint. To break loose from this constraint, MIMO technology \cite{sun2020mimo,liao2018fast,ren2019mimo} is used in the state-of-the-art high-resolution automotive radar to increase the resolution of the system when the number of physical antennas is fixed. For a 3-transmitter/4-receiver MIMO radar system, one may construct a virtual array of 12 elements using only 7 physical antennas. To further improve the resolution, sparse array designs \cite{vaidyanathan2010sparse,liao2013direction,hu2013doa} are often coupled with MIMO virtual array approach to increase the effective array aperture while reducing the hardware cost and mutual coupling among antennas \cite{liao2012adaptive,liao2012doa,liao2011direction}. Fig.~\ref{fig:virtual_array} depicts a sparse array of 12 virtual elements spanning an aperture of the size of 28 elements. Sub-$1^{\circ}$ angular resolution has been achieved in practical 4D imaging radar systems by employing MIMO sparse array approach.

One main challenge in sparse array design is to deal with angular ambiguity and sidelobes. Moreover, in highly dynamic scenarios, angle estimation has to be conducted based on single snapshot measurement \cite{hacker2010single}. Many conventional angle estimation methods fail to operate under such conditions \cite{liao2010subspace,li1993performance,weber2009analysis}. In this paper, we focus on high-resolution direction-of-arrival (DoA) estimation \cite{liao2018array} with MIMO sparse array using only a single snapshot in automotive radar applications.   

In a typical automotive frequency-modulated continuous-wave (FMCW) radar processing chain, the targets are first separated in range and Doppler domains. Due to the wide bandwidth feature of current automotive radars, the number of targets that falls in the same range-Doppler bin is small \cite{sun2020mimo}. Therefore,  high-resolution results in angle space can be obtained by solving a sparse signal recovery (SSR) problem \cite{DOAcs}. Note that sparse signal recovery provides a framework to effectively handle sparse signals encountered in many signal processing applications, such as spectral analysis \cite{spectral}, sparse channel estimation \cite{rice2019sparse}, and multiuser detection \cite{Usercs}. It is also a common problem found in machine learning, such as feature selection \cite{feature}, subspace clustering \cite{subspace}, and sparse representation \cite{face}.

The general linear model for the sparse signal recovery problem \cite{baraniuk2007} can be presented as:
\begin{equation}
    \mathbf{y}=\mathbf{Ac}+\boldsymbol{\epsilon}
    \label{eq:linear_model}
\end{equation}
where $\mathbf{y}$ represents the $M$ observed measurements, 
$\mathbf{c}$ is the $N$ unknown parameters with $N\gg M$,
$\mathbf{A}$ is the basis matrix, and $\boldsymbol{\epsilon}$ is the noise vector. Equation (\ref{eq:linear_model}) is under-determined and generally has infinite numbers of solutions.
Various approaches have been proposed to find the sparsest solution of (\ref{eq:linear_model}), including but not limited to, Greedy-based approaches \cite{JA07,cosamp,JA04}, Convex relaxation approaches \cite{LASSO1996,BP2001,JA06}, Iterative reweighted approaches \cite{RL1,FOCUSS,irlp,RL12}, and Bayesian linear regression approaches \cite{wipf04,SJ08,SD10,MD98,sblrvm}.

Among the aforementioned approaches, Bayesian linear regression approaches are superior in scalability and interpretability. There are different ways to classify these approaches. Here, we consider two related frameworks: the maximum a posteriori (MAP) and the hierarchical frameworks. 
The main difference between these two frameworks lies in how the sparsity-inducing prior is used.
In the latter, the prior is used in a hierarchical manner, while in the former, it is used directly. Two particular approaches are of particular interest here. The first approach is
the Cauchy-Gaussian (CG) approach in \cite{MD98}, which also belongs to the iterative reweighted algorithms from the implementation perspective \cite{blunt2011robust}. It uses the MAP framework with the sparsity-inducing Cauchy prior 
\begin{equation}
    p(c_i;\gamma)=\frac{1}{\pi\gamma(1+\frac{c_i^2}{\gamma^2})}
    \label{eq:cauchy_prior}
\end{equation}
where $\gamma$ in (\ref{eq:cauchy_prior}) is the scale parameter. CG is very effective in suppressing spurious targets if its hyper-parameters are chosen properly. However, the problem of solving the posterior in CG is intractable, so systematic strategies for updating the CG hyper-parameters are missing. Without such strategies, the optimal hyper-parameters can only be found by trial and error in a suboptimal way.

The second approach is the sparse Bayesian learning (SBL) approach which has been widely used in radar signal processing \cite{chen2018off,nannuru2018sparse,wang2016sparse}. SBL uses the hierarchical framework with the conditional Gaussian prior \cite{sblrvm}
of the unknown vector $\mathbf{c}$:
\begin{equation}
    p(\mathbf{c}|\pmb{\tau})=\frac{1}{\sqrt{(2\pi)^N|\mathbf{\Sigma}^{-1}|}}
    \text{exp}\big(-\frac{1}{2}\mathbf{c}^T\mathbf{\Sigma c}\big)
    \label{eq:sbl_prior}
\end{equation}
where the diagonal matrix $\mathbf{\Sigma} \equiv \text{diag}(\pmb{\tau})$ is the inverse of the covariance matrix of prior $\mathbf{c}$ conditional on
the $N \times 1$ hyper-parameter $\pmb{\tau}$. Compared to the aforementioned algorithms, SBL excels in sparse signal recovery performance with uniform linear array and sufficient signal-to-noise ratio (SNR), and its parameters are self-tuned. However, recent studies have highlighted several areas in which SBL can be improved, particularly when sparse array measurement is used as input. Firstly, SBL involves $N$ hyper-parameters in $\pmb{\tau}$ to define $N$ Gaussian distributions corresponding to the $N$ elements of $\mathbf{c}$. As elements of $\pmb{\tau}$ approach infinity at various speeds during the iterations, the condition number of the matrix to be inverted in SBL increases. Coupled with limited machine precision in real-world applications (e.g., single-precision computing environments in most automotive radar processors), this results in numerical errors and instability\cite{sblrvm}.
Thus, the iteration process has to be stopped before numerical error takes place. Secondly, the noise parameter $\sigma_n$ estimated by SBL converges incorrectly resulting in suboptimal solution \cite{sigman2018,DP07,Wipfphd} and negatively impacting the sparse recovery performance in practice \cite{zhilin2011temporal}. Lastly, because of the above-mentioned reasons, SBL tends to produce more spurious solutions \cite{sblrvm,nannuru2017multi,nannuru2017sparse}, which impairs its usefulness when high dynamic range is required. This issue becomes increasingly pronounced as the sparse array becomes sparser. It is essential to highlight that the primary focus of this work is on applications that employ sparse array measurements as input. Numerous SBL applications also employ a uniform linear array as input. For instance, reference \cite{liao2022map} offers a valuable example of the SBL application that utilizes a uniform linear array.

In this paper, a sparse signal recovery algorithm, Bayesian linear regression with Cauchy prior (BLRC), is proposed to improve the performance of CG and SBL, specifically when sparse array measurements serve as input data. Like CG, BLRC has only one hyper-parameter $\gamma$ for all $N$ prior distributions for the $N$ elements of the unknown $\mathbf{c}$. 
Unlike CG, we use Laplace approximation in BLRC to approximate the intractable posterior locally
so that the hyper-parameters can be learned from the observed data through the proposed approximate expectation maximization (AEM) scheme. 
In this sense, BLRC is a significant improvement over the CG approach 
because BLRC provides a systematic updating scheme for the hyper-parameters which is absent from the CG approach. 

BLRC can also be considered as a significant improvement over SBL in several aspects. Like SBL, BLRC is a Bayesian linear regression approach.
Both BLRC and SBL have the same iterative updating steps for the solution and hyper-parameters. However, there are some key differences between BLRC and SBL.
Primarily, only two hyper-parameters need to be handled in BLRC, while the number of hyper-parameters to be handled is large for SBL. This is because BLRC uses a long-tailed but proper Cauchy distribution that requires only one scale parameter to characterize all $N$ elements of the unknown
$\mathbf{c}$. 
On the contrary, SBL uses the Gaussian distribution as the intermediate conditional prior where its true prior is improper \cite{sblrvm}.
Accordingly, SBL needs N variances to characterize the $N$ elements of the unknown
$\mathbf{c}$.
As the number of unknowns is typically large, the number of hyper-parameters for SBL is usually large.
Therefore, BLRC is superior to SBL in latent space compactness, which is a major advantage of BLRC over SBL.

Using randomly placed Sparse Array (SPA) and Coprime Array (CPA), it is shown in this paper by extensive numerical simulations that BLRC has the following advantages over SBL:
numerical robustness, spurious targets suppression, target resolution,
system flexibility, noise tolerance, and noise variance estimation. This is originated from the fact that one scale parameter $\gamma$ is sufficient to describe the Cauchy prior of BLRC while $N$ hyper-parameters $\{ \tau_i \} $ are required to describe the conditional Gaussian priors of SBL. Furthermore, the performance of BLRC and SBL are compared using sparse automotive radar image recovery examples. 
Using physical optics (PO) approximation, MIMO radar signals are generated by our automotive imaging radar simulator. It is shown that BLRC outperforms SBL by producing clearer radar images with better resolution.

The rest of this paper is organized as follows. In Section II, we give a summary of the Bayesian linear regression model, CG approach, and SBL approach. In Section III, we use AEM to develop the proposed BLRC approach and discuss its computational efficiency. In Section IV, comparisons under variational interpretation are provided to explain the improvement of BLRC over CG. In Section V, comparisons based on the iterative reweighted $l_2$ (IR-$l_2$) interpretation are provided to explain the superior performance of BLRC over SBL. In Section VI, numerical analyses including initial values of hyper-parameters, numerical stability, convergent properties, spurious targets suppression, noise variance estimation, resolution and sensitivities study are provided. In Section VII, we introduce the automotive image radar signal processing application and our automotive radar signal simulator. The direction of arrival (DoA) estimation result using coprime array is shown to demonstrate the superiority of our proposed BLRC approach. At last, the conclusion is made in Section VIII. 

\section{Sparse Bayesian Linear Regression Approaches}
In this section, we introduce the MAP framework and the hierarchical framework for sparse Bayesian linear regression approaches. Then we briefly summarize the Cauchy Gaussian (CG) algorithm, which uses the MAP framework with Cauchy prior, and the well-known Sparse Bayesian Learning (SBL), which uses the hierarchical framework.

\subsection{Sparse Bayesian Linear Regression}
Using the general linear model for the sparse signal recovery in (\ref{eq:linear_model}) and assuming the $M$ noise elements of 
$\boldsymbol{\epsilon}$ are independent zero-mean Gaussian random variables with the same variance $\sigma_n^2$, the likelihood function of the observation $\mathbf{y}$ is:
\begin{equation}
    p(\mathbf{y}|\mathbf{c},{\sigma_n}) = (\frac{1}{\sqrt{2\pi \sigma_n^2}})^{M}
    \text{exp}\big({-\frac{||\mathbf{y-Ac}||^2}{{2\sigma_n^2}}}\big)
    \label{eq:likelihood}
\end{equation}

Using the MAP framework, we are to find the mode of the following posterior distribution:
\begin{equation}
\begin{aligned}
   \mathbf{c} = \arg \max_{\mathbf{c}}~    p(\mathbf{c}| \mathbf{y})
   = \arg \max_{\mathbf{c}}~{\ln p(\mathbf{y}|\mathbf{c})p(\mathbf{c})}
    \label{eq:MAP}
\end{aligned}
\end{equation}
The prior distribution $p(\mathbf{c})$ is a sparsity-inducing prior which can typically be written in the following general form:
\begin{equation}
    p(\mathbf{c})\propto \exp{\big(-\frac{1}{2}\sum_{i=1}^Ng(c_i)\big)}
\end{equation}
where $g(c_i)=h(c_i^2)$, and it has been shown that the choice of a concave and nondecreasing $h(c_i)$ in $[0,\infty)$ could lead to a sparse solution of $\mathbf{c}$ \cite{Wipf11}. Then (\ref{eq:MAP}) leads to the canonical regularized optimization problem:
\begin{equation}
    \mathbf{c} =\arg \min_{\mathbf{c}}~{||\mathbf{y-Ac}||^2+\sigma_n^2 \sum_{i=1}^Ng(c_i)}
    \label{eq:generalMAP}
\end{equation}
For example, $h(c_i)=\sqrt{c_i}$ results in the Laplacian prior which is a well-known sparsity-inducing prior, and the corresponding $g(c_i)=|c_i|$ leads to the well-known regularized optimization problem with $l_1$ norm.

The hierarchical framework is proposed to form the full Bayesian inference by introducing extra hyper-parameters (e.g., $\pmb{\tau}$ in SBL \cite{sblrvm}). Then, not only $\mathbf{c}$, but also the hyper-parameter $\pmb{\tau}$ and noise variance $\sigma_n^2$ are all inferred from observation $\mathbf{y}$:
\begin{equation}
\begin{aligned}
   \{\Tilde{\mathbf{c}},\Tilde{\pmb{\tau}},\Tilde{\sigma}_n \} & = \arg \max_{\mathbf{c},\pmb{\tau},\sigma_n}~p(\mathbf{c},\pmb{\tau},\sigma_n| \mathbf{y})\\
   &= \arg \max_{\mathbf{c},\pmb{\tau},\sigma_n}~ {p(\mathbf{c}|\mathbf{y},\pmb{\tau},\sigma_n)p(\pmb{\tau},\sigma_n|\mathbf{y})}
   \label{eq:hier0}
\end{aligned}
\end{equation}
To solve the above inference problem, several algorithms are proposed, such as the evidence maximization (also known as type-II maximum likelihood) \cite{sblrvm}, expectation maximization (EM) \cite{EM}, and variational Bayes (VB) approximation \cite{VB}. 

Basically, (\ref{eq:hier0}) can be solved by the following iterative approach.
Firstly, the hyper-parameter $\Tilde{\pmb{\tau}}$ and $\Tilde{\sigma}_n$ are learned based on observation $\mathbf{y}$ using the following optimization:
\begin{equation}
\begin{aligned}
     \{\Tilde{\pmb{\tau}},\Tilde{\sigma}_n\}&=\arg \max_{\pmb{\tau},\sigma_n}~p(\pmb{\tau},\sigma_n|\mathbf{y})\\
     &= \arg \max_{\pmb{\tau}, \sigma_n }
   p(\mathbf{y}|\pmb{\tau}, \sigma_n)\\
    &=\arg \max_{\pmb{\tau},\sigma_n}~ \int p(\mathbf{y}|\mathbf{c},\sigma_n)p(\mathbf{c}|\pmb{\tau})d\mathbf{c}
    \label{eq:hier1}
\end{aligned}
\end{equation}
with the assumption that $\pmb{\tau}$ and $\sigma_n$ are ``improper" hyper-parameters with  flat distributions  ($p(\pmb{\tau})\propto 1$ and $p(\sigma_n) \propto 1$) or non-informative Jeffreys distributions (uniform under logarithmic scale). 

Secondly, the estimation of $\Tilde{\mathbf{c}}^{(k+1)}$ of $\mathbf{c}$ for the $(k+1)^{th}$ iteration is obtained from maximizing the posterior distribution:
\begin{equation}
\begin{aligned}
   \Tilde{\mathbf{c}}^{(k+1)}= \arg \max_{\mathbf{c}}~ p(\mathbf{c}|\mathbf{y},\pmb{\tau}(\Tilde{\mathbf{c}}^{(k)}),
   \Tilde{\sigma}_n(\mathbf{c}^{(k)}))
   \label{eq:hier2}
\end{aligned}
\end{equation}
If the posterior distribution is Gaussian, the solution of (\ref{eq:hier2}) is the mean of the posterior distribution.
Using (\ref{eq:hier1}) and (\ref{eq:hier2}), the $\mathbf{c}$ and $\{\pmb{\tau},\sigma_n \}$ are updated in an iterative manner until a convergence criterion is satisfied.

Note that the ``true" prior distribution $p(\mathbf{c})$ can be obtained from the parameterized prior $p(\mathbf{c}|\pmb{\tau})$ in the hierarchical framework:
\begin{equation}
    p(\mathbf{c})=\int p(\mathbf{c}|\pmb{\tau})p(\pmb{\tau})d\pmb{\tau}
    \label{eq:true_prior}
\end{equation}
It is shown that the ``true" prior $p(\mathbf{c})$ is actually a sparsity-inducing prior, and this kind of representation is also known as scale mixtures \cite{giri2016}. 

\subsection{Cauchy Gaussian (CG) Approach}

The CG approach \cite{MD98} uses the MAP framework with the Cauchy prior given in (\ref{eq:cauchy_prior}).
Assuming the scale parameter $\gamma$ and the noise standard deviation $\sigma_n$ are given (which are denoted as $\hat{\gamma}$ and $\hat{\sigma}_n$, respectively), the MAP estimator for $\mathbf{c}$ is to minimize the following cost function.  
\begin{equation}
 J_{cg}(\mathbf{c}) 
    =||\mathbf{y-Ac}||^2+h_{cg}(\mathbf{c})
\label{eq:J_cg}
\end{equation}
with the log-sum regularized term
\begin{equation}
\begin{aligned}
    h_{cg}(\mathbf{c}) 
    ={\hat{\sigma}_n^2}\sum_{i=1}^{N}2\ln(1+\frac{c_i^2}{\hat{\gamma}^2})
\end{aligned}
\label{eq:joint_pdf1}
\end{equation}
Omitting the derivations, $\mathbf{c}$ is computed iteratively using steps described in Algorithm 1.

\begin{algorithm}
\DontPrintSemicolon
  \SetAlgoLined
  \KwInput{\ $\mathbf{y}$, \quad $\mathbf{A}$,\quad $\hat{\sigma}_n$,\quad $\hat{\gamma}$, \quad $K$}
  \KwOutput{\ $\hat{\mathbf{c}}$}
  \textbf{Initialization:}$\  \hat{\mathbf{c}}^{(0)} $\;
\For{$k = 1:K$}{
 $\hat{\mathbf{Q}}^{(k)}=\textrm{diag}\Big(1+\frac{(\hat{c}^{(k-1)}_1)^2}{\hat{\gamma}^2},
 ~1+\frac{(\hat{c}^{(k-1)}_2)^2}{\hat{\gamma}^2},\cdots,
 ~1+\frac{(\hat{c}^{(k-1)}_{N})^2}{\hat{\gamma}^2}~\Big)^{-1}$\;
 $\hat{\mathbf{c}}^{(k)} = [\frac{2\hat{\sigma}_n^2}{\hat{\gamma}^2} \hat{\mathbf{Q}}^{(k)}+\mathbf{A}^T\mathbf{A}]^{-1}\mathbf{A}^T\mathbf{y}$\;
  \text{if} $\hat{\mathbf{c}}^{(k)}$ \text{converges}, $k=K$ and $\hat{\mathbf{c}}^{(K)}=\hat{\mathbf{c}}^{(k)}$
}
\Return{$\hat{\mathbf{c}}=\hat{\mathbf{c}}^{(K)}$}\;
\caption{Cauchy Gaussian (CG) Approach}
\end{algorithm}

\subsection{Sparse Bayesian Learning (SBL)}
In sparse Bayesian Learning (SBL), the prior is represented in a hierarchical framework. The conditional prior distribution of the unknown vector $\mathbf{c}$ is Gaussian as defined in (\ref{eq:sbl_prior}).
Although there are different explanations and optimization algorithms for SBL (e.g., evidence maximization, expectation maximization, and variational Bayes), their updating strategies for the sparse vector $\mathbf{c}$ are similar.

For convenience, let the $i^{th}$ element of $\pmb{\tau}$, $\tau_i=1/\sigma_i^2$, where $\sigma_i^2$ is the variance of $c_i$, the $i^{th}$ element of $\mathbf{c}$.
Additionally, let ${\mathbf{\Gamma}}$ be the covariance matrix of the posterior of $\mathbf{c}$ conditional on $\mathbf{y}$.
Denote the estimate of $\mathbf{c}$, $\pmb{\tau}$, $\mathbf{\Sigma}$, $\mathbf{\Gamma}$ and $\sigma_n$
as $\tilde{\mathbf{c}}$, $\tilde{\pmb{\tau}}$, $\tilde{\mathbf{\Sigma}}$, $\tilde{\mathbf{\Gamma}}$ and $\tilde{\sigma}_n$, respectively.
The typical implementation of SBL as shown in \cite{SJ08} and \cite{sblrvm} for finding $\tilde{\mathbf{c}}$ is summarized in Algorithm 2.

\begin{algorithm}
\DontPrintSemicolon
\SetAlgoLined
\KwInput{\ $\mathbf{y}$, \quad $\mathbf{A}$, \quad $K$, \quad \textrm{size}($\mathbf{A}$)=[\it{M}, \it{N}\rm{]}}
\KwOutput{\ $\tilde{\mathbf{c}}$}
\textbf{Initialization:}$\ \tilde{\sigma}_n^{(0)} \gets 0.1,\quad \tilde{\pmb{\tau}}^{(0)} \gets \mathbf{1}, \quad k \gets 0$\;
\For{$k = 1:K$}{
 $\tilde{\mathbf{\Sigma}}^{(k)}=\textrm{diag}\Big(\tilde{\tau}_1^{(k-1)},~\tilde{\tau}_2^{(k-1)},\cdots,~\tilde{\tau}_{N}^{(k-1)}~\Big)$\;
 $\tilde{\mathbf{\Gamma}}^{(k)}=(\frac{1}{(\tilde{\sigma}_n^{(k-1)})^2}
 \mathbf{A}^T\mathbf{A}+\tilde{\mathbf{\Sigma}}^{(k)})^{-1}$\;
 $\tilde{\mathbf{c}}^{(k)}=\frac{1}{(\tilde{\sigma}_n^{(k-1)})^2}
 \tilde{\mathbf{\Gamma}}^{(k)}\mathbf{A}^T\mathbf{y}$\;
 $(\tilde{\sigma}_n^{(k)})^2=\frac{|| \mathbf{y}-\mathbf{A}\tilde{\mathbf{c}}^{(k)}||^2}{M-\text{Tr}(\mathbf{I}-\tilde{\mathbf{\Gamma}}^{(k)}\tilde{\mathbf{\Sigma}}^{(k)})}$\;
 \For{$i = 1:N$}{
 $ \tilde{\tau}_i^{(k)} = \frac{1-\tilde{\tau}_i^{(k-1)}
 \tilde{\mathbf{\Gamma}}_{ii}^{(k)}}
 {(\tilde{c}_i^{(k)})^2}$\;
 }
 \text{if} $\tilde{\mathbf{c}}^{(k)}$ \text{converges}, $k=K$ and $\tilde{\mathbf{c}}^{(K)}=\tilde{\mathbf{c}}^{(k)}$\;
}
\Return{$\tilde{\mathbf{c}}=\tilde{\mathbf{c}}^{(K)}$ }\;
\caption{Sparse Bayesian Learning (SBL)}
\end{algorithm}

\section{Bayesian Linear Regression with Cauchy Prior (BLRC)}

When the parameters $\sigma_n$ and $\gamma$ are given for the CG approach, one can directly compute 
$\mathbf{c}$ and $\mathbf{Q}$ iteratively as shown in Algorithm 1. 
However, $\sigma_n$ and $\gamma$ are usually unknown for the CG approach and are typically found by trial and error under practical conditions. 
In this section, 
using the Laplace approximation to obtain a local Gaussian approximation of the posterior distribution of the random vector $\mathbf{c}$,
we present a novel Approximate Expectation Maximization (AEM) algorithm to update these parameters automatically. Benefiting from the general convergence property of EM algorithm, the proposed algorithm is to converge to a local minimum. 
The combination of the CG approach for estimating $\mathbf{c}$ and $\mathbf{Q}$ and the AEM approach for updating $\sigma_n$ and $\gamma$ is named as the Bayesian Linear Regression with Cauchy Prior (BLRC) approach.

\subsection{Expectation Maximization (EM) Formulation}
To obtain the maximum likelihood estimation of parameter $\{\gamma, \sigma_n \}$, 
we need to maximize 
$\ln \{p(\mathbf{y};\gamma, \sigma_n)\}$:
\begin{equation}
   \{\hat{\gamma}, \hat{\sigma}_n \} = \arg\max_{\{\gamma, \sigma_n \}}
   \ln \{ p(\mathbf{y};\gamma, \sigma_n) \}
    \label{eq:MAP_BLRC2}
\end{equation}

Since $p(\mathbf{y};\gamma,\sigma_n)$ does not have a closed-form expression with the Cauchy prior model,
we propose to maximize its lower bound (i.e., the generalized EM formulation \cite{bishop2006pattern}). Then the estimator for $\{\gamma,\sigma_n\}$ in the $k^{th}$ iteration can be expressed as
\begin{equation}
\begin{aligned}
    &\{\hat{\gamma}^{(k)},\hat{\sigma}_n^{(k)}\}
    =\arg\max_{\gamma,\sigma_n}
    E_{\mathbf{c|y}}[\ln p(\mathbf{y,c};\gamma,\sigma_n)]
    \label{eq:EM}
\end{aligned}
\end{equation}
where $E_{\mathbf{c|y}}$ is the expectation with respect to the random variable ${\mathbf{c}}$ given the obervation ${\mathbf{y}}$ and the parameters from the previous iteration $\{\gamma^{(k-1)},\sigma_n^{(k-1)}\}$

For convenience, the parameter $\gamma$ in Cauchy prior distribution is mapped to a new parameter $\tau$ with $\gamma=\frac{1}{\sqrt{\tau}}$ for further mathematical manipulations.
Noting that $p(\mathbf{y,c};\tau,\sigma_n)=p(\mathbf{y}|\mathbf{c};\sigma_n)p(\mathbf{c};\tau)$ and using the likelihood distribution of observation $p(\mathbf{y}|\mathbf{c};\sigma_n)$ in (\ref{eq:likelihood}) and the Cauchy prior distribution $p(\mathbf{c};\tau)$ in (\ref{eq:cauchy_prior}), the expectation in (\ref{eq:EM}) can be expressed as
\begin{equation}
    \begin{aligned}
    &E_{\mathbf{c|y}}[\ln p(\mathbf{y,c};\tau,\sigma_n)]\\
    &= -\frac{M}{2}\ln (2\pi \sigma_n^2)-\frac{1}{2\sigma_n^2}E_{\mathbf{c|y}}[||\mathbf{y-Ac} ||^2]\\
    &~~~-N\ln (\frac{\pi}{\sqrt{\tau}})-\sum_{i=1}^{N}E_{\mathbf{c|y}}[\ln(1+\tau c_i^2)]
    \label{eq:expectation}
    \end{aligned}
\end{equation}
where the computation of the original expectation can be carried out by computing the two new easier expectations.
Define the mean $E_{\mathbf{c|y}}[\mathbf{c}]$ as $\mathbf{\hat{c}}$ and the covariance $E_{\mathbf{c|y}}[(\mathbf{c}-\mathbf{\hat{c}})(\mathbf{c}-\mathbf{\hat{c}})^T]$ as $\mathbf{\hat{\Gamma}}$. The first expectation in (\ref{eq:expectation}) can be further simplified following Appendix A:
\begin{equation}
    \begin{aligned}
    E_{\mathbf{c|y}}[||\mathbf{y-Ac} ||^2]
    =||\mathbf{y-A\hat{\mathbf{c}}}  ||^2+ \text{Tr}(\mathbf{A}^T\mathbf{A}\hat{\mathbf{\Gamma}})
    \label{eq:exp_2}
    \end{aligned}
\end{equation}

Regarding the second expectation in (\ref{eq:expectation}), it is necessary to simplify  
$\ln(1+\tau c_i^2)$ in order to get a closed-form expression.
Consider the second-order Taylor series expansion of $\ln(1+\tau c_i^2)$ at $\tau c_i^2=E_{\mathbf{c|y}}[\tau c_i^2]$:
\begin{equation}
    \begin{aligned}
    \ln(1+\tau c_i^2)&\approx \ln(1+E_{\mathbf{c|y}}[\tau c_i^2])+\frac{\tau c_i^2-E_{\mathbf{c|y}}[\tau c_i^2]}{(1+E_{\mathbf{c|y}}[\tau c_i^2])\cdot 1!}\\
    &-\frac{(\tau c_i^2-E_{\mathbf{c|y}}[\tau c_i^2])^2}{(1+E_{\mathbf{c|y}}[\tau c_i^2])^2\cdot 2!}
    \label{eq:ln}
    \end{aligned}
\end{equation}
Taking expectation on (\ref{eq:ln}), we have:
\begin{equation}
    E_{\mathbf{c|y}}[\ln(1+\tau c_i^2)]
    \approx\ln(1+\tau E_{\mathbf{c|y}}[c_i^2])-\frac{\tau^2 Var_{\mathbf{c|y}}[c_i^2]}
    {2(1+\tau E_{\mathbf{c|y}}[c_i^2])^2}
    \label{eq:exp_3}
\end{equation}
For convenience, define $\eta_i=c_i^2$, $\hat{\eta}_i=E_{\mathbf{c|y}}[\eta_i]$
and $\hat{\xi}_i=Var_{\mathbf{c|y}}[\eta_i]$. Then (\ref{eq:exp_3}) becomes
\begin{equation}
    E_{\mathbf{c|y}}[\ln(1+\tau \eta_i)]
    \approx \ln(1+\tau \hat{\eta}_i )-\frac{\tau^2 \hat{\xi}_i}
    {2(1+\tau \hat{\eta}_i)^2}
    \label{eq:exp_4}
\end{equation}

\subsection{Approximate Expectation}
Closed-form expressions of the mean and variance of ${\mathbf{c}}$ (i.e., $\hat{\mathbf{c}}$ and $\hat{\mathbf{\Gamma}}$, respectively) in (\ref{eq:exp_2})
and the mean and variance of $\eta_i$ (i.e., $\hat{\eta}_i$ and $\hat{\xi}_i$, respectively) in (\ref{eq:exp_4})
are difficult to derive since there is no explicit expression of the posterior distribution $p(\mathbf{c}|\mathbf{y};\tau,\sigma_n)$.

Although Cauchy distribution is non-log-concave, the Hessian matrix of the posterior can be shown to be negative-definite everywhere when being solved in an iterative reweighted manner. Thus, the iteratively solved posterior $p(\mathbf{c}|\mathbf{y})$ is  log-concave and unimodal. Then, the Laplace approximation, a general methodology that approximates a probability density function locally in terms of a Gaussian distribution, can be employed to approximate $p(\mathbf{c}|\mathbf{y};\tau,\sigma_n)$.

%Since Cauchy prior in (\ref{eq:cauchy_prior}) is the distribution of the magnitude $|{\mathbf{c}}|$, we will at first treat ${\mathbf{c}}$ as a real random vector for convenience. After obtaining the probability density function by making the Laplace approximation, we will then extend ${\mathbf{c}}$ to be a complex random vector.

Performing Taylor series expansion on the previously defined cost function in (\ref{eq:J_cg})
around the mode $\hat{\mathbf{c}}$, we obtain:
\begin{equation}
    \begin{aligned}
    J_{cg}(\mathbf{c})
    %&\approx
    %J(\hat{\mathbf{c}})+(\mathbf{c}-\hat{\mathbf{c}})^T\frac{\partial J(\mathbf{c})}{\partial \mathbf{c}}\Big|_{\mathbf{c}=\hat{\mathbf{c}}}\\
    %&~~~~~~~~~+\frac{1}{2}(\mathbf{c}-\hat{\mathbf{c}})^T\frac{\partial^2J(\mathbf{c})}{\partial\mathbf{c}^2}\Big|_{\mathbf{c}=\hat{\mathbf{c}}}(\mathbf{c}-\hat{\mathbf{c}})\\
    &\approx J_{cg}(\hat{\mathbf{c}})+\frac{1}{2}(\mathbf{c}-\hat{\mathbf{c}})^T \hat{\mathbf{\Gamma}}^{-1}(\mathbf{c}-\hat{\mathbf{c}})
    \end{aligned}
    \label{eq:J}
\end{equation}
where $\hat{\mathbf{c}}$ is given in line 4 of Algorithm 1 and
\begin{equation}
    \hat{\mathbf{\Gamma}}^{-1}=\nabla_{\mathbf{c}} \nabla_{\mathbf{c}} J_{cg}(\mathbf{c})\Big|_{\mathbf{c}=\hat{\mathbf{c}}}
    \label{eq:Gamma}
\end{equation}
%Noting $\frac{\partial J(\mathbf{c})}{\partial \mathbf{c}}\Big|_{\mathbf{c}=\hat{\mathbf{c}}}=0$,
Then, using the Laplace approximation, we obtain the local Gaussian approximation of the posterior distribution of the random vector $\mathbf{c}$ from (\ref{eq:J}):
\begin{equation}
    \begin{aligned}
    p(\mathbf{c}|\mathbf{y};\tau,\sigma_n)&\propto p(\mathbf{y}|\mathbf{c};\sigma_n)p(\mathbf{c};\tau)\\
    &\simeq \frac{1}{(2\pi)^{\frac{N}{2}}
    |\hat{\mathbf{\Gamma}}|^{\frac{1}{2}}}\exp(-\frac{1}{2}(\mathbf{c}-\hat{\mathbf{c}})^T \hat{\mathbf{\Gamma}}^{-1}(\mathbf{c}-\hat{\mathbf{c}}))
    \end{aligned}
    \label{eq:real}
\end{equation}
Note that the covariance matrix $\hat{\mathbf{\Gamma}}$ in (\ref{eq:real}) can be obtained from (\ref{eq:Gamma}).
Since $\hat{\mathbf{Q}}$ in line 3 of Algorithm 1 is a function of
$\hat{\mathbf{c}}$,
we expressed $\hat{\mathbf{\Gamma}}^{(k)}$ of the $k^{th}$ iteration in terms of the following iterative formula:
\begin{equation}
    \begin{aligned}
    \hat{\mathbf{\Gamma}}^{(k)}
    =\Bigg[\frac{1}{\sigma_n^2}\mathbf{A}^T\mathbf{A}
    +2\tau\hat{\mathbf{Q}}^{(k)} \Bigg]^{-1}
    \end{aligned}
    \label{eq:post_Gammai}
\end{equation}
So we can rewrite the approximate posterior mean under Cauchy prior in line 4 of Algorithm 1 as:
\begin{equation}
    \hat{\mathbf{c}}^{(k)} = \frac{1}{\sigma_n^2}\hat{\mathbf{\Gamma}}^{(k)}\mathbf{A}^T\mathbf{y}
    \label{eq:post_mui}
\end{equation}

Regarding the mean and variance of $\eta_i$ (i.e., $\hat{\eta}_i$ and $\hat{\xi}_i$) in (\ref{eq:exp_4}), we can use the identity:
\[
    \begin{aligned}
    \hat{\eta_i}&=E_{\mathbf{c|y}}[c_i^2]
    =Var_{\mathbf{c|y}}[c_i]+(E_{\mathbf{c|y}}[c_i])^2\\
    \hat{\xi}_i&=Var_{\mathbf{c|y}}[c_i^2]
    =E_{\mathbf{c|y}}[c_i^4]-(E_{\mathbf{c|y}}[c_i^2])^2\\
    \end{aligned}
\]
Note that $Var_{\mathbf{c|y}}[c_i]=\hat{\mathbf{\Gamma}}_{ii}$, $(E_{\mathbf{c|y}}[c_i])^2=\hat{c}_i^2$
%Since the posterior distribution is approximated as Gaussian distribution, 
%we can get $E_{\mathbf{c|y}}[|c_i|^4]$ from its characteristic function $\Phi_{c_i}(\omega)$:
%\begin{equation}
%    E[|c_i|^k]=\frac{1}{j^k}\frac{\pa%rtial^k\Phi_{c_i}(\omega)}{\partial %\omega^k} \Bigg|_{\omega=0}
%\end{equation}
%where $\Phi_{c_i}(\omega)=e^{\hat{c}_i\omega+\frac{\omega^2\hat{\Gamma}_{ii}^2}{2}}$.
%Then we get:
,and 
%\begin{equation}
   $ E[c_i^4]= \hat{c}_i^4
    +6\hat{c}_i^2\hat{\mathbf{\Gamma}}_{ii}+3\hat{\mathbf{\Gamma}}_{ii}^2$. 
%\end{equation}
Finally we obtain
\begin{equation}
    \begin{aligned}
    \hat{\eta_i}&=\hat{\mathbf{\Gamma}}_{ii}
    +\hat{c}_i^2\\
    \hat{\xi}_i&=
    4\hat{c}_i^2\hat{\mathbf{\Gamma}}_{ii}+2\hat{\mathbf{\Gamma}}_{ii}^2 
    \label{eq:etaxi}
    \end{aligned}
\end{equation}

\subsection{Maximization}

Setting the derivatives of (\ref{eq:expectation}) with respect to $\sigma_n^2$ and $\tau$ to zeros, one can obtain the maximization results for (\ref{eq:EM}).
Firstly, taking the derivative of (\ref{eq:expectation}) with respect to $\sigma_n^2$ and utilizing (\ref{eq:exp_2}), 
we have
\begin{equation}
    \begin{aligned}
    \frac{\partial E_{\mathbf{c|y}}[\ln p(\mathbf{y,c})] }{\partial (\sigma_n^2)}=-\frac{M}{2\sigma_n^2}+
    \frac{||\mathbf{y-A\hat{\mathbf{c}}}  ||^2+ \text{Tr}(\mathbf{A}^T\mathbf{A}\hat{\mathbf{\Gamma}})}{2(\sigma_n^2)^2} 
    \label{eq:max1}
    \end{aligned}
\end{equation}

From (\ref{eq:max1}), the estimate of $\sigma_n^2$, denoted as $\hat{\sigma}_n^2$, can be obtained as shown in line 9 of Algorithm 3.
Noting that the $\hat{\mathbf{c}}$ and $\hat{\mathbf{\Gamma}}$ in (\ref{eq:max1}) are functions of $\hat{{\sigma}}_n^2$, so $\hat{{\sigma}}_n^2$ has to be solved iteratively.

Secondly, taking derivative of (\ref{eq:expectation}) with respect to $\tau$ and utilizing (\ref{eq:exp_3}), 
we have 
\begin{equation}
    \begin{aligned}
    &\frac{\partial E_{\mathbf{c|y}}[\ln p(\mathbf{y,c})] }{\partial \tau}=\frac{N}{2\tau}-\sum_{i=1}^{N}\frac{\hat{\eta_i}}{1+\tau \hat{\eta_i}}+\sum_{i=1}^{N}\frac{\tau \hat{\xi}_i}{(1+\tau \hat{\eta_i})^3}\\
       \end{aligned}
       \label{eq:exp_5}
\end{equation}
Thus, we can solve iteratively for the update of $\tau$, denoted as $\hat{\tau}$, as shown in line 8 of Algorithm 3. In the proposed BLRC algorithm, the update of $\{\hat{\mathbf{c}},  \hat{\mathbf{\Gamma}} \}$ and the update of 
$\{ \hat{\sigma}_n^2 , \hat{\tau}\}$ are carried out iteratively, which is summarized in Algorithm 3.

\begin{algorithm}
\DontPrintSemicolon
\SetAlgoLined 
  \KwInput{\ $\mathbf{y}$, \quad $\mathbf{A}$, \quad $K$, \quad \textrm{size}($\mathbf{A}$)=[\it{M},\it{N}\rm{]}}
  \KwOutput{\ $\hat{\mathbf{c}}$}
  \textbf{Initialization:}$\ \hat{\sigma}_n^{(0)} \gets 0.1,\quad \hat{\gamma}^{(0)} \gets 1,\quad \hat{\mathbf{c}}^{(0)} $\;
\For{$k = 1:K$}{
 $
  \begin{aligned}
  \hat{\mathbf{Q}}^{(k)}&=\textrm{diag}\Big(1+\frac{(\hat{c}^{(k-1)}_1)^{2}}{(\hat{\gamma}^{(k-1)})^2},\cdots,1+\frac{(\hat{c}^{(k-1)}_{N})^{2}}{(\hat{\gamma}^{(k-1)})^2}~\Big)^{-1}
  \end{aligned}
 $\;
 $ 
 \hat{\mathbf{\Gamma}}^{(k)}
    =\Bigg[\frac{1}{(\hat{\sigma}_n^2)^{(k-1)}}\mathbf{A}^T\mathbf{A}
    +\frac{2}{(\hat{\gamma}^{(k-1)})^2}\hat{\mathbf{Q}}^{(k)} \Bigg]^{-1}
 $\;
 $
 \hat{\mathbf{c}}^{(k)}=
 \frac{1}{(\hat{\sigma}_n^2)^{(k-1)}}
 \hat{\mathbf{\Gamma}}^{(k)}\mathbf{A}^T\mathbf{y}
  $\;
$
\hat{\eta}_i^{(k)}=\hat{\mathbf{\Gamma}}_{ii}^{(k)}+(\hat{c}_i^{(k)})^2
$\;
$ 
\hat{\xi}_i^{(k)}=
    4(\hat{c}_i^{(k)})^2\hat{\mathbf{\Gamma}}_{ii}^{(k)}+2(\hat{\mathbf{\Gamma}}_{ii}^{(k)})^2
    $\;
 $     
(\hat{\gamma}^{(k)})^2=\frac{2}{N}
 \Big\{
 {\sum_{i=1}^{N}\Big[\frac{\hat{\eta_i}^{(k)}}{1+\frac{\hat{\eta_i}^{(k)}}{(\hat{\gamma}^{(k-1)})^2} }-\frac{\frac{1}{(\hat{\gamma}^{(k-1)})^2} \hat{\xi}_i^{(k)}}{(1+\frac{1}{(\hat{\gamma}^{(k-1)})^2} \hat{\eta_i}^{(k)})^3}  } \Big]\Big\}
 $\;
  $    (\hat{\sigma}_n^2)^{(k)} 
  =\frac{1}M \Big\{ ||\mathbf{y-A\hat{\mathbf{c}}}^{(k)}  ||^2+ \text{Tr}(\mathbf{A}^T\mathbf{A}\hat{\mathbf{\Gamma}}^{(k)})\Big\}
  $\;
 \text{if} $\hat{\mathbf{c}}^{(k)}$ \text{converges}, $k=K$ and $\hat{\mathbf{c}}^{(K)}=\hat{\mathbf{c}}^{(k)}$\;
}
\Return{$\hat{\mathbf{c}}=\hat{\mathbf{c}}^{(K)}$}\;
\caption{Bayesian Linear Regression with Cauchy Prior (BLRC)}
\end{algorithm}

\subsection{Computational Efficiency and Pruning}
Computational complexities of the three Bayesian regression approaches (BLRC, CG and SBL) are more or less of the same order. This is because matrix inversion is the most computationally intense step (see step 4 of Algorithms 1, 2 and 3) and the matrices to be inverted are of the same size, $N\times N$. where $N$ is usually large.

\begin{figure}[!t]
\centering
  \includegraphics[width=3.3in]{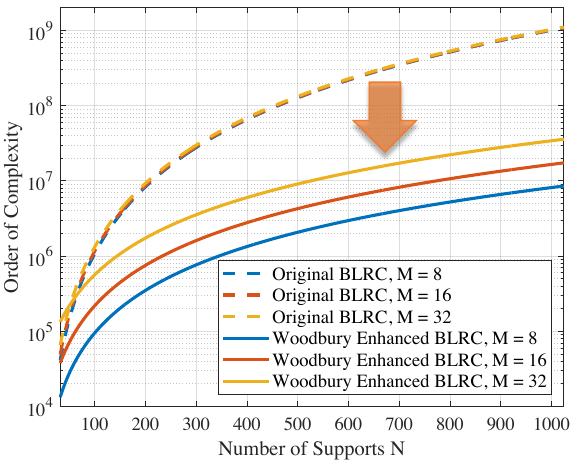}
  \caption{Computational Efficiency Enhancement via using Woodbury Matrix Identity}
  \label{fig:complexity}
\end{figure}

In order to be used in real-time processing applications such as automotive radar, speed-up methods are required for CG, SBL, and BLRC to improve their computational efficiency. Outlined below are three standard strategies.
Firstly, the Woodbury matrix identity can be employed to reduce the computational complexity of inverting the $N\times N$ matrices because 
$\hat{Q}$ and $\tilde{\Sigma}$ are diagonal matrices and
the system matrix $\mathbf{A}$ is $M\times N$, $M \ll N$. 
As shown in Fig. \ref{fig:complexity}, computation speed has been improved by one to two orders of magnitude. Secondly,
further increase of the computational efficiency is to use pruning to reduce the effective size of $\mathbf{A}$ in each iteration. If a pruning threshold $t_p^{(k)}$ for the $k^{th}$ iteration is determined, the elements of ${\mathbf{c}^{(k-1)}}$ with magnitudes smaller than $t_p^{(k)}$ are marked.
These elements and their corresponding columns in $\mathbf{A}$ are excluded.
Consequently, the size of the matrix to be inverted is reduced in the $k^{th}$ iteration. 
Thirdly, further improvement on computational efficiency can be made by using efficient matrix inversion algorithms such as Cholesky decomposition for the matrix inversion.

In such a manner, the computation is not only more efficient but also more robust against numerical issues. However, pruning leads to sub-optimal results \cite{sblrvm}. Moreover, selecting the pruning threshold $t_p^{(k)}$ itself is not a trivial problem. The larger the pruning threshold is, the more numerically efficient the three Bayesian regression approaches will be, but with an increased risk of discarding weak targets. If the pruning threshold is too small, it has little effect on the computation efficiency. Thus, for fully understanding the convergent property of BLRC,  we do not use pruning in numerical analyses in this paper.

\section{Comparison under Variational Interpretation}
In this section, we firstly compare the true prior used in SBL and the Cauchy prior used in both CG and BLRC. Secondly, in order to directly compare different algorithms in the same $\mathbf{c}$-space, the variational bounds of the priors in SBL and BLRC are shown. Note that the core of deriving variational bound is to represent a convex function using its dual form \cite{jordan99}, and the variational bound of SBL has been derived in \cite{Wipf11} which will be omitted here. Finally, cost functions of $l_p$, CG, SBL and BLRC in $\mathbf{c}$-space are compared. It is shown that, like SBL, BLRC has the capability to reduce the number of local minimums which explains the superior performance of BLRC over CG. 

\subsection{True Prior Comparison}

The intrinsic true prior used in SBL can be obtained by integrating 
$p(c_i|\tau_i)p(\tau_i)$
with respect to the hyper-parameter $\tau_i$. Using (\ref{eq:true_prior}),
the integration result can be represented as a special case of the Student-t distribution $St(c_i;0,\sqrt{\frac{b}{a}},2a)$ \cite{sblrvm}:
\begin{equation}
\begin{aligned}
     p(c_i;a,b)&=\int p(c_i|\tau_i)p(\tau_i)d\tau_i\\
     &=\int \mathcal{N}(c_i|0,\tau_i^{-1})Gamma(\tau_i|a,b)d\tau_i\\
     &=\frac{b^a\mathit{\Gamma}(a+\frac{1}{2})}{(2\pi)^{\frac{1}{2}}\mathit{\Gamma}(a)}(b+\frac{c_i^2}{2})^{-(a+\frac{1}{2})}
     \label{eq:pc_student}
\end{aligned}
\end{equation}
where $a,b$ are the parameters of Gamma distribution. 
Specifically, the parameters are set to $a=1$ and $b=0$ to get an improper uniform hyper-prior \cite{RVM} or set to $a=b=0$ to get the Jeffreys prior \cite{sblrvm}. Then the true prior degenerates from the Student-t distribution to an improper prior $p(c_i)\propto \frac{1}{|c_i|^3}$ or $p(c_i)\propto \frac{1}{|c_i|}$, respectively. Note that they are improper priors because the areas under these probability density curves cannot be equal to one in the absence of properly defined scaling factors. Although it is hard to directly use these improper priors for Bayesian inference, the MAP estimation adopting these improper priors leads to the IR-$l_1$ algorithm. 

The non-standard Cauchy prior $Cauchy(c_i;0,\gamma)$ in (\ref{eq:cauchy_prior}) for both CG and BLRC is 
also a special case of non-standard Student-t distribution $St(c_i;0,\gamma,\nu)$ with $\nu=1$:
\begin{equation}
    St(c_i;0,\gamma,\nu)=\frac{\mathit{\Gamma}(\frac{\nu+1}{2})}{\mathit{\Gamma}(\frac{\nu}{2})}(\frac{1}{\pi\nu\gamma^2})^{\frac{1}{2}}[1+\frac{c_i^2}{\nu\gamma^2}]^{-\frac{\nu+1}{2}}
    \label{eq:stu}
\end{equation}
where $\gamma$ is the scale parameter and $\nu$ is the degree of freedom.
Note that (\ref{eq:stu}) becomes (\ref{eq:pc_student}) if $\nu=2a$ and $\gamma^2=\frac{b}{a}$.
And the Cauchy prior used in CG and BLRC can be obtained by setting $a=\frac{1}{2}$ which is between 0 and 1. 

Unlike the improper prior for SBL, the obtained Cauchy prior for BLRC is a proper distribution. 
Compared with the Student-t distribution which has a parameter $a$ embedded in the gamma function, Cauchy distribution does not have the parameter $a$ and is much easier to be used for Bayesian inference. 
Moreover, since the $\pmb{\tau}$ has been integrated out in (\ref{eq:pc_student}), there is no need to estimate the large $N\times 1$ vector $\pmb{\tau}$ for Bayesian inference as in the case of SBL. Instead, in the BLRC approach, one needs to estimate only the scalar parameter $\gamma$ which is related to the parameter $b$ in (\ref{eq:pc_student}).

\subsection{Conjugate Dual and Variational Bound}
The cost functions of algorithms with MAP framework (e.g., CG) lie in the $\mathbf{c}$-space, whereas the cost functions of algorithms with hierarchical framework (e.g., SBL and BLRC) are not directly represented in $\mathbf{c}$-space. Fortunately, as shown in \cite{Wipf11}, it provides a way to derive the cost function of algorithms with hierarchical framework in $\mathbf{c}$-space using conjugate dual and variational bound.

From the Student-t prior of a scalar $c_i$ in (\ref{eq:pc_student}) and omitting the subscript $i$ for convenience, we have
\begin{equation}
\begin{aligned}
    \ln p(c)= -(a+\frac{1}{2})\ln(b+\frac{c^2}{2})+\ln\frac{b^a\mathit{\Gamma}(a+\frac{1}{2})}{(2\pi)^{\frac{1}{2}}\mathit{\Gamma}(a)}
\end{aligned}
\end{equation}
Define the function $f(x)$:
\begin{equation}
    f(x)=-(a+\frac{1}{2})\ln(b+\frac{x}{2})+\ln\frac{b^a\mathit{\Gamma}(a+\frac{1}{2})}{(2\pi)^{\frac{1}{2}}\mathit{\Gamma}(a)}
    \label{eq:fx}
\end{equation}
where $\ln p(c)=f(c^2)$ and $f(x)$ is convex.
Also define the conjugate function of $f(x)$ as:
\begin{equation}
    f^*(\xi) = \sup_{x}~\big\{\frac{-\xi}{2}x-f(x) \big\},~\xi>0
        \label{eq:fx_1}
\end{equation}
where we use $\frac{-\xi}{2}x$ instead of $\xi x$ for notational convenience.
Substituting (\ref{eq:fx}) into (\ref{eq:fx_1}), 
it can be shown that
\begin{equation}
    x^*=\frac{2a+1}{\xi}-2b
    \label{eq:fx_2}
\end{equation}
is the solution to $f^*(\xi)$ in (\ref{eq:fx_1}).
Let $x=x^*$ in (\ref{eq:fx_2}), 
the conjugate function in (\ref{eq:fx_1}) becomes
\begin{equation}
\begin{aligned}
    f^*(\xi)&=-(a+\frac{1}{2})+\xi b+(a+\frac{1}{2})\ln \frac{2a+1}{2\xi}\\
    &-\ln\frac{b^a\mathit{\Gamma}(a+\frac{1}{2})}{(2\pi)^{\frac{1}{2}}\mathit{\Gamma}(a)},~~\xi>0,~a>0,~b>0
        \label{eq:fx_3}
\end{aligned}
\end{equation}
Using (\ref{eq:fx_1}) and (\ref{eq:fx_3}), we then derive the lower bound of $f(x)$:
\begin{equation}
\begin{aligned}
     f(x)&\ge \frac{-\xi}{2}x-f^*(\xi)\\
     &=-\frac{\xi x}{2}+a+\frac{1}{2}-\xi b-(a+\frac{1}{2})\ln \frac{2a+1}{2\xi}\\
     &+\ln\frac{b^a\mathit{\Gamma}(a+\frac{1}{2})}{(2\pi)^{\frac{1}{2}}\mathit{\Gamma}(a)}
             \label{eq:fx_4}
\end{aligned}
\end{equation}
Since $p(c)=e^{f(c^2)}$, we can use (\ref{eq:fx_4}) to get the lower bound of the Student-t prior $p(c)$ as
\begin{equation}
    \begin{aligned}
         p(c) \ge
      \frac{1}{\sqrt{2\pi \xi^{-1}}}e^{-\frac{c^2}{2\xi^{-1}}}\cdot \phi(\xi,a,b),~~\forall \xi>0,~a>0,~b>0
      \label{eq:convexbound}
    \end{aligned}
\end{equation}
where
\begin{equation}
    \begin{aligned}
        \phi(\xi,a,b)=\xi^ae^{-b\xi}\cdot \frac{b^a\mathit{\Gamma}(a+\frac{1}{2})}{\mathit{\Gamma}(a)}\cdot (\frac{2a+1}{2e})^{-(a+\frac{1}{2})}
    \end{aligned}
    \label{eq:phi}
\end{equation}

If $a\rightarrow 0$ and $b\rightarrow 0$, it is easy to see that a Jeffrey's non-informative prior is obtained. The evidence maximization can be applied and the cost function of SBL in $\mathbf{c}$-space can be derived as shown in \cite{Wipf11} which is also provided in Appendix B:
\begin{equation}
    \min_{\mathbf{c}} J_{sbl}(\mathbf{c}),~~J_{sbl}(\mathbf{c})=|| \mathbf{y}-\mathbf{Ac}||^2+h_{sbl}(\mathbf{c})
    \label{eq:L_sbl}
\end{equation}
where the minimization with respect to $\pmb{\tau}$ and $\sigma_n$ is done as following (in deriving the regularization $h_{sbl}(\mathbf{c})$
to encourage a sparse solution):
\begin{equation}
    h_{sbl}(\mathbf{c})=\min_{\pmb{\tau},\sigma_n}~\sigma_n^2 \Big(\mathbf{c}^T\mathbf{\Sigma}\mathbf{c}+ \ln |\sigma_n^2\mathbf{I}+\mathbf{A}\mathbf{\Sigma}^{-1}\mathbf{A}^T| \Big)
    \label{eq:h_sbl}
\end{equation}

\subsection{BLRC Cost Function in $\mathbf{c}$-space}

By setting $a=\frac{1}{2}$ and $b=\frac{\gamma^2}{2}$ in (\ref{eq:convexbound}),
a lower bound of the Cauchy prior $p(c)$ is shown to be:
\begin{equation}
\begin{aligned}
     p(c) \ge
      \frac{1}{\sqrt{2\pi \xi^{-1}}}e^{-\frac{c^2}{2\xi^{-1}}}\cdot \phi(\xi,\gamma)
      = \mathcal{N}({0},\xi^{-1})\phi(\xi,\gamma)
    \label{eq:p_c}
\end{aligned}
\end{equation}
where $\phi(\xi,\gamma)=\gamma e \sqrt{\frac{\xi}{2\pi}}e^{-\frac{\xi\gamma^2}{2}},~~\xi>0,~\gamma>0$. 
Note that the bound in (\ref{eq:p_c}) is tight when
\begin{equation}
    \xi = \frac{2}{c^2+\gamma^2}.
    \label{eq:tight}
\end{equation}
Replacing $c$ and $\xi$ by $c_i$ and $\xi_i$, respectively, in both (\ref{eq:p_c}) and (\ref{eq:tight}),
we can find $p(c_i)$ from (\ref{eq:p_c}). 
Then, using $p(c_i),~\forall i$, we can extend the lower bound in (\ref{eq:p_c}) of a scalar $c$ to a lower bound of a vector $\mathbf{c}$:
\begin{equation}
\begin{aligned}
     p(\mathbf{c}; \gamma)=\prod_{i=1}^N p(c_i)
     \ge \mathcal{N}(\mathbf{0},\mathbf{\Theta}^{-1})\prod_{i=1}^N \phi(\xi_i,\gamma)
     \label{eq:pc_lowerbound}
\end{aligned}
\end{equation}
where 
\begin{equation}
    \mathbf{\Theta}=\text{diag}(\xi_1,\xi_2,...,\xi_N)
    \label{eq:Theta}
\end{equation}
In (\ref{eq:pc_lowerbound}), the dependence on $\gamma$ is explicitly shown. Given that $\xi_i$ and $c_i$ are related according to (\ref{eq:tight}), the lower bound in (\ref{eq:pc_lowerbound}) is tight.

With the likelihood function in (\ref{eq:likelihood}) and the lower bound of Cauchy prior in (\ref{eq:pc_lowerbound}) at hand, a lower bound of the evidence distribution $p(\mathbf{y}; \gamma, \sigma_n)$ can be expressed as the following integral:
\begin{equation}
    \begin{aligned}
         p(\mathbf{y}; \gamma, \sigma_n)&= \int p(\mathbf{y}|\mathbf{c}; \sigma_n )p(\mathbf{c}; \gamma)d\mathbf{c} \\
         &\ge \int \mathcal{N}(\mathbf{Ac},\sigma_n^2\mathbf{I})\mathcal{N}(\mathbf{0},\mathbf{\Theta}^{-1}) 
         \prod_{i=1}^N \phi(\xi_i,\gamma)d\mathbf{c}
    \end{aligned}
    \label{eq:p_y_gamma}
\end{equation}
where the lower bound is tight.
Unfortunately, we cannot obtain a closed-form expression of the above integral.
In Section III, we have already used AEM to derive an alternative lower bound of $p(\mathbf{y}; \gamma, \sigma_n)$ (see (\ref{eq:EM}) and subsequent equations).

Here, we will approximate the integral in (\ref{eq:p_y_gamma}) based on the iterative procedure of BLRC.
Note that the estimate of $c_i$ for the $(k-1)^{th}$ iteration, $c_i^{(k-1)}$, is a known constant, 
so the following corresponding parameters
\begin{equation}
\begin{aligned}
    \xi_i^{(k-1)} &= \frac{2}{(c_i^{(k-1)})^2+\gamma^2}\\
    \mathbf{\Theta}^{(k-1)}&=\text{diag}(\xi_1^{(k-1)},\xi_2^{(k-1)},...,\xi_N^{(k-1)})
    \label{eq:tight_iter}
\end{aligned} 
\end{equation}
are also known constants.
Then, $\phi(\xi_i^{(k-1)},\gamma)), ~\forall i,$ derived from the $(k-1)^{th}$ iteration are independent of the integration variable $\mathbf{c}$ and 
can be brought out of the integral in (\ref{eq:p_y_gamma}).
Thus, (\ref{eq:p_y_gamma}) becomes
\begin{equation}
    \begin{aligned}
         p(\mathbf{y}; \gamma,\sigma_n) \ge \mathcal{N}(\mathbf{0},\mathbf{\Theta}_y^{-1})\cdot \prod_{i=1}^N \phi(\xi_i^{(k-1)},\gamma)
         \label{eq:lowerbound_p_y}
    \end{aligned}
\end{equation}
with
\begin{equation}
    \begin{aligned}
         \mathcal{N}(\mathbf{0},\mathbf{\Theta}_y^{-1})= \int \mathcal{N}(\mathbf{Ac},\sigma_n^2\mathbf{I})\mathcal{N}(\mathbf{0},(\mathbf{\Theta}^{(k-1)})^{-1})d\mathbf{c}
         \label{eq:lowerbound_p_y_1}
    \end{aligned}
\end{equation}
where 
\begin{equation}
  \mathbf{\Theta}_y^{-1}=\sigma_n^2\mathbf{I}+\mathbf{A}(\mathbf{\Theta}^{(k-1)})^{-1}\mathbf{A}^T
  \label{eq:Theta_y}
\end{equation}
In (\ref{eq:lowerbound_p_y}), the lower bound may not be tight because of the iterative approximation. 

Note that maximizing $\ln p(\mathbf{y}; \gamma,\sigma_n)$ is equivalent to minimizing the negative logarithm of the lower bound of $p(\mathbf{y}; \gamma,\sigma_n)$. 
The cost function to be minimized for the $k^{th}$ iteration for BLRC is then defined as
\begin{equation}
    \begin{aligned} L_{blrc}=\ln |\mathbf{\Theta}_y^{-1}|+\mathbf{y}^T\mathbf{\Theta}_y\mathbf{y}-\sum_{i=1}^N \ln \phi(\xi_i^{(k-1)},\gamma)
    \end{aligned}
        \label{eq:L_blrc1}
\end{equation}
Recall that $c_i^{(k-1)}$ is defined by the previous iteration, so $L_{blrc}$ is not shown to be an explicit function of $\mathbf{c}$ in (\ref{eq:L_blrc1}).
However, using the same procedure given in Appendix C, it can be shown that
\begin{equation}
    \begin{aligned}
        \mathbf{y}^T\mathbf{\Theta}_y\mathbf{y}=\min_{\mathbf{c}}~ \Big(\frac{1}{\sigma_n^2}||\mathbf{y}-\mathbf{Ac}||^2+\mathbf{c}^T\mathbf{\Theta}^{(k-1)}\mathbf{c}\Big)
    \end{aligned}
    \label{eq:L_blrc2}
\end{equation}
Substituting (\ref{eq:L_blrc2}) into (\ref{eq:L_blrc1}), $L_{blrc}$ can then be viewed as a function of $\mathbf{c}$ as shown below:
\begin{equation}
    \begin{aligned}
    L_{blrc}(\mathbf{c})   &=\min_{\mathbf{c}}~\Big( \frac{1}{\sigma_n^2}|| \mathbf{y}-\mathbf{Ac}||^2+\mathbf{c}^T\mathbf{\Theta}^{(k-1)}\mathbf{c}\Big)\\
    &+ \ln |\mathbf{\Theta}_y^{-1}|-\sum_{i=1}^N \ln \phi(\xi_i^{(k-1)},\gamma)
    \label{eq:L_blrc3}
    \end{aligned}
\end{equation}
Note that the last two terms in (\ref{eq:L_blrc3}) do not depend on the unknown $\mathbf{c}$ for the current iteration.
Substituting (\ref{eq:Theta_y}) into (\ref{eq:L_blrc3}), minimizing $L_{blrc}$ with respect to $\gamma$ and $\sigma_n$ can be rewritten in $\mathbf{c}$-space as:
\begin{equation}
    \min_{\mathbf{c}} J_{blrc}(\mathbf{c}),~~J_{blrc}(\mathbf{c})= || \mathbf{y}-\mathbf{Ac}||^2+h_{blrc}(\mathbf{c})
    \label{eq:L_blrc}
\end{equation}
where the minimization with respect to $\gamma$ and $\sigma_n$ is carried out in executing the regularization $h_{blrc}(\mathbf{c})$:
\begin{equation}
\begin{aligned}
    h_{blrc}(\mathbf{c})&=\min_{\gamma, \sigma_n}~ \sigma_n^2
    \Big(-\sum_{i=1}^N \ln \phi(\xi_i^{(k-1)},\gamma) \\
    &    \mathbf{c}^T\mathbf{\Theta}^{(k-1)}\mathbf{c}+ \ln |\sigma_n^2\mathbf{I}+\mathbf{A}(\mathbf{\Theta}^{(k-1)})^{-1}\mathbf{A}^T|\Big)
    \label{eq:blrc_cspace}
\end{aligned}
\end{equation}
Here, the regularization $h_{blrc}(\mathbf{c})$ is to encourage a sparse solution.

\subsection{Comparison in $\mathbf{c}$-space}

Consider the following maximally sparse signal recovery problem:
\begin{equation}
    \min_{\mathbf{c}}~||\mathbf{c}||_0, ~~s.t.  ~\mathbf{y}=\mathbf{Ac}
    \label{eq:max_sparse_0}
\end{equation}
Here, the $l_0$ norm, $||\mathbf{c}||_0$, represents the number of non-zero elements in vector $\mathbf{c}$. Although the cost function in (\ref{eq:max_sparse_0}) is ideal for many applications, particularly when exact reconstruction \cite{shen2013exact} is desired, finding its global minimum is an NP-hard problem.

Given that $||\mathbf{c}||_0\equiv \lim_{p\rightarrow 0}\sum_i |c_i|^p$, we employ the $l_p$ norm (i.e., $||\mathbf{c}||_p^p$ with $0< p< 1$) as a benchmark in this section: 
\begin{equation}
    \min_{\mathbf{c}}~||\mathbf{c}||_p^p, ~~s.t.  ~\mathbf{y}=\mathbf{Ac}
    \label{eq:max_sparse}
\end{equation}
We then compare optimization landscapes of the cost functions in (\ref{eq:max_sparse}), (\ref{eq:J_cg}), (\ref{eq:L_sbl}), and (\ref{eq:L_blrc})
corresponding to $l_p$, CG, SBL, and BLRC, respectively. An effective optimization landscape is expected to possess the same global minimum as (\ref{eq:max_sparse_0}), while exhibiting fewer local minimums. 

\begin{figure}[!t]
\centering
  \includegraphics[width=3.3in]{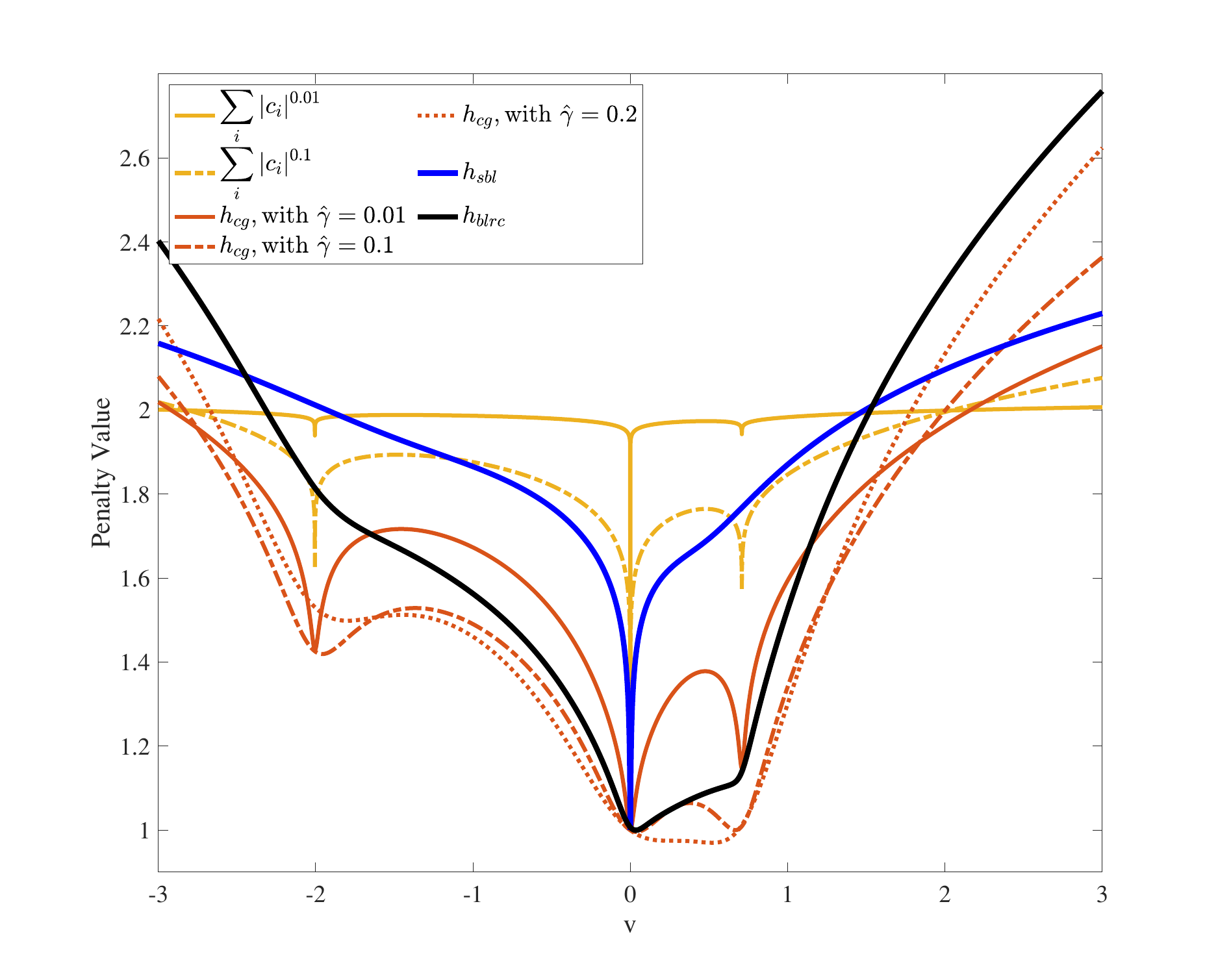}
  \caption{Normalized penalty values of (\ref{eq:max_sparse}), (\ref{eq:J_cg}), (\ref{eq:L_sbl}), and (\ref{eq:L_blrc}) with respect to the parameter $\rm{v}$ in (\ref{eq:minimum_profile}).}
  \label{fig:localmin}
\end{figure}

In order to visualize the differences of optimization landscapes defined in (\ref{eq:max_sparse}), (\ref{eq:J_cg}), (\ref{eq:L_sbl}), and (\ref{eq:L_blrc}), 
consider the example shown in \cite{Wipf11} where $N = M+1$. 
Thus, the null-space of $\mathbf{A}$,
denoted by $\mathbf{a}_{null}$, has only one dimension. 
If $\mathbf{c}_{op}$ is the optimum solution achieving 
the global minimum of (\ref{eq:max_sparse_0}),
$\mathbf{c}_{op}$ shows maximum sparsity and
\begin{equation}
   \mathbf{y=Ac}=\mathbf{Ac}_{op},~ \mathbf{c} = \mathbf{c}_{op}+\mathrm{v}\cdot\mathbf{a}_{null}
   \label{eq:minimum_profile}
\end{equation}
where $\mathrm{v}$ is an arbitrary constant.
Choosing an $\mathbf{A}$ where $\mathbf{c}_{op}$ is known,
we can then plot the regularization terms in (\ref{eq:max_sparse}), (\ref{eq:J_cg}), (\ref{eq:L_sbl}), and (\ref{eq:L_blrc}), with respect to $\mathbf{c}$ by changing the scalar $\rm{v}$, to view the optimization landscape (see Fig.~\ref{fig:localmin}). Since the constraint $\mathbf{y=Ac}$ is always satisfied with different $\mathrm{v}$, the optimization landscape of regularization term is also the optimization landscape of the corresponding optimization problem. Note that we scale the regularization terms such that all the regularization terms equal to $1$ at $\rm{v}=0$.

Since $l_p$ is close to the $l_0$ if $p$ is small. In Fig.~\ref{fig:localmin}, the $p=0.01$ curve is similar to that for the $l_0$ norm, where the large pit at $\rm{v}=0$ denotes the global minimum. Unfortunately, there exist two other pits at $\rm{v}=-2$ and $\rm{v}=0.71$, which represent two local minimums. As expected, the large pit at $\rm{v}=0$ is widened while maintaining its position as global minimum when $p$ increases to $0.1$.
The two other pits at $\rm{v}=-2$ and $\rm{v}=0.71$ are also widened with the increase in $p$. Nonetheless, the existence of these local minimums persists with various $p$ value in $(0,1)$, which complicates the search for the global minimum.

As shown in Fig.~\ref{fig:localmin}, 
it is remarkable that $h_{cg}$ in (\ref{eq:joint_pdf1}) with a small $\hat{\gamma}$ value for the CG approach behaves like $l_p$ norm with a small $p$ value. 
As $\hat{\gamma}$ increases, all pits get smoother.
Thus, in order to get a more sparse solution, we need to choose a smaller $\hat{\gamma}$ for the CG approach.
However, the CG iteration is more likely to be trapped in one of the local minimums when $\hat{\gamma}$ is small.
On the other hand,
if we choose a large $\hat{\gamma}$ for iteration, we may avoid being trapped in a local minimum. However, the solution is no longer accurate because 
the global minimum may shift to a different location.
For example,
when $\hat{\gamma}=0.2$, the global minimum shifts from $\rm{v}=0$ to $\rm{v}=0.52$ as shown in Fig.~\ref{fig:localmin}.
Therefore, it is unclear how to choose $\hat{\gamma}$ in general for the CG approach.

To plot the penalty $h_{sbl}(\mathbf{c})$ in (\ref{eq:h_sbl}) for SBL, the $\mathbf{c}$ is still controlled by the scalar $\rm{v}$ using (\ref{eq:minimum_profile}). However, the hyper-parameters $\pmb{\tau}$ and $\sigma_n$ within $h_{sbl}(\mathbf{c})$ cannot be predetermined because they depend on the value of $\mathbf{c}$. 
Thus for a given $\rm{v}$, we firstly compute $\mathbf{c}$ using (\ref{eq:minimum_profile}). Then, use line 6 and 8 in Algorithm 2 to compute $\sigma_n$ and $\pmb{\tau}$ iteratively until minimum $h_{sbl}(\mathbf{c})$ is obtained. In Fig.~\ref{fig:localmin}, the SBL curve shows the correct global minimum but no local minimums. 
Note that the $h_{sbl}$ curve has been shown in \cite{Wipf11} to produce fewer local minimums than $l_p$ approach as well.

Similar to SBL, the hyper-parameters $\mathbf{\gamma}$ and $\sigma_n$ of BLRC depend on the value of $\mathbf{c}$.
Thus, to plot the penalty $h_{blrc}(\mathbf{c})$ in (\ref{eq:blrc_cspace}),
we firstly compute $\mathbf{c}$ using (\ref{eq:minimum_profile}) for a given $\rm{v}$. 
Then, noting that $\gamma=\frac{1}{\sqrt{\tau}}$,
use line 8 and 9 in Algorithm 3 to compute $\sigma_n$ and ${\tau}$ iteratively
until minimum $h_{blrc}(\mathbf{c})$ is reached. In Fig.~\ref{fig:localmin}, BLRC presents the correct global minimum without any local minimums, which indicates that it has similar capability as SBL to reduce the number of local minimums.
This explains the superior performance of BLRC over CG seen in numerical examples in Section VI.

\section{Comparisons under IR-$l_2$ Interpretations}

In this section, we will compare BLRC, SBL, and CG based on the IR-$l_2$ interpretation which will be used to explain the superior performances of BLRC over SBL in Section VI.

\subsection{IR-$l_2$ Formulation of CG and BLRC}

Since both CG and BLRC use the Cauchy prior in the MAP framework, 
they have exactly the same steps for updating $\mathbf{c}$ 
(see step 4 in Algorithm 1 and steps 4 and 5 of Algorithm 3). These updating formulas are derived from solving the log-sum regularized optimization problem 
\begin{equation}
    {\mathbf{c}} = \arg \min_{\mathbf{c}} 
    ||\mathbf{y-Ac}||^2+ {{\sigma}_n^2}\sum_{i=1}^{N}2\ln({c_i^2}+{\hat{\gamma}^2}).
\label{eq:joint_pdf}
\end{equation}
It is well-known that the log-sum penalty encourages a sparse solution. We typically use the following IR-$l_2$ approach to minimize this objective function in (\ref{eq:joint_pdf}): 
\begin{equation}
    {\mathbf{c}}^{(k+1)}=\arg\min_{\mathbf{c}}~\{||\mathbf{y-Ac}||^2_2+\sigma_n^2 \sum_{i=1}^N {w}_i^{(k)}c_i^2\}
    \label{eq:ir_l2}
\end{equation}
where the weight for the $k^{th}$ iteration is
\begin{equation}
    {w}_i^{(k)} = \frac{2}{(\hat{\gamma}^{(k)})^2+(\hat{c}_i^{(k)})^2}
    \label{eq:cg_weight}
\end{equation}
Note that $\{\hat{c}_i^{(k)} \}$ in (\ref{eq:cg_weight}) are obtained from the $k^{th}$ iteration and are considered as constants in (\ref{eq:ir_l2}).
The noise variance $\sigma_n^2$ in (\ref{eq:ir_l2}) and the scale parameter $\hat{\gamma}^{(k)}$ in (\ref{eq:cg_weight}) are constants with respect to $k$ in CG,
but are updated at each $k$ in BLRC.

\subsection{IR-$l_2$ Formulation of SBL}

For the $\mathbf{c}$ updating step in SBL, it is remarkable that line 5 of Algorithm 2 coincides with the solution of the following log-sum regularized optimization problem:
\begin{equation}
\begin{aligned}
    \min_{\mathbf{c}}~&||\mathbf{y-Ac}||^2_2+\sigma_n^2\sum_{i=1}^{N}\ln(c_i^2+\tilde{\mathbf{\Gamma}}_{ii})
    \label{eq:sbl_IR}
\end{aligned}
\end{equation}
The IR-$l_2$ updating scheme for solving the above optimization problem is given in (\ref{eq:ir_l2})
where the weight 
\begin{equation}
    w_i^{(k)} = \frac{1}{\tilde{\mathbf{\Gamma}}^{(k)}_{ii}+(\tilde{c}_i^{(k)})^2}
    \label{eq:sbl_weight}
\end{equation}
Let $\tau_i^{(k)}= w_i^{(k)}$. It can be shown that (\ref{eq:sbl_weight}) is equivalent to the
updated hyper-parameter $\tilde{\pmb{\tau}}$ at the $k^{th}$ iteration in step 8 of Algorithm 2. 
Thus, in SBL, we actually ``learn" the hyper-parameter $\tau_i^{(k)}$ using the IR-$l_2$ updating scheme.

\subsection{The IR-$l_2$ Point of View}

From the IR-$l_2$ point of view, SBL, BLRC and CG solve the same equation (\ref{eq:ir_l2}) except the following difference.
CG and BLRC use a common factor $(\hat{\gamma}^{(k)})^2$ for all weights in (\ref{eq:cg_weight}).
However, SBL needs different factors $\tilde{\mathbf{\Gamma}}_{ii}$ for different weights in (\ref{eq:sbl_weight}).

\begin{figure}[!t]
  \centering
  \includegraphics[width=3.3in]{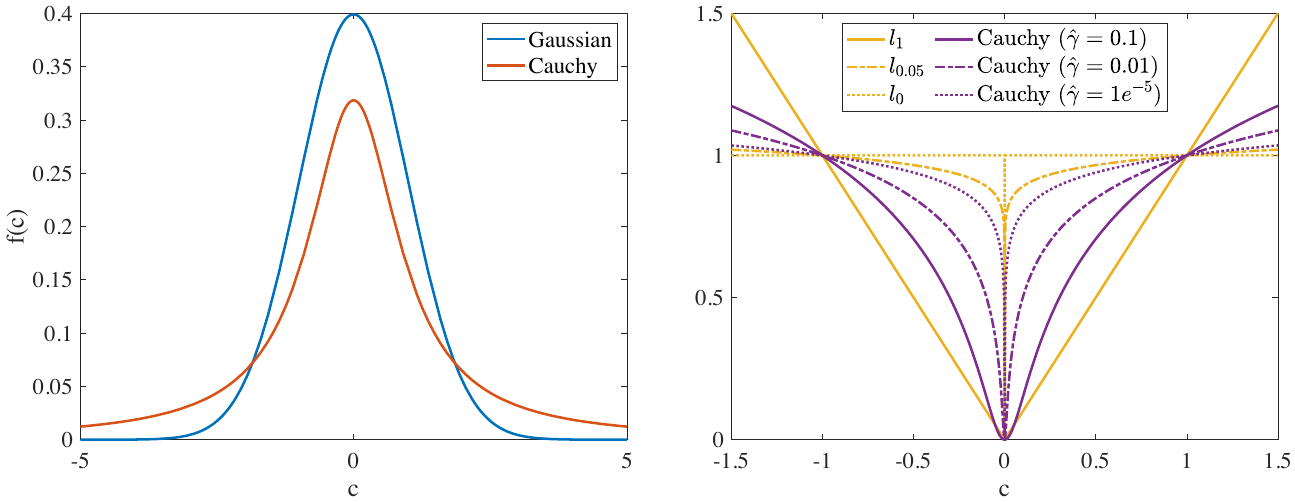}
  \caption{(Left) Comparison of Gaussian ($\sigma=1$) and Cauchy ($\gamma=1$) Distributions; (Right) Illustration of Regularization Functions for CG and BLRC}
  \label{fig:GC}
\end{figure}

As shown in the left subplot of Fig.~\ref{fig:GC}, Cauchy in (\ref{eq:cauchy_prior}) is a long-tailed distribution.
When the parameter $\gamma$ approaches zero, most of the probability mass will concentrate at zero, which implies most $c_i$'s will have high probabilities to be zero.
However, no matter how small $\gamma$ is, the mean and variance of Cauchy distribution are still undefined.
This implies that a small number of $c_i$'s can be non-zeros.
Thus, a single universal parameter $\gamma$ of Cauchy prior can be used to model all $c_i$'s as seen in (\ref{eq:cg_weight}). 

This is not the case for the conditional Gaussian prior. As shown in the left subplot of Fig.~\ref{fig:GC}, 
Gaussian in (\ref{eq:sbl_prior}) is not a long-tailed distribution.
When a particular $c_i$ is near zero,
the variance $\sigma_i^2$  (i.e., $1/\tau_i$) of the corresponding zero-mean Gaussian distribution must approach zero,
and vice versa.
Similarly, when a particular $c_i$ is non-zero,
the variance $\sigma_i^2$ of the corresponding zero-mean Gaussian distribution must not be near zero,
and vice versa.
Therefore, with Gaussian prior, $N$ ${\tilde{\mathbf{\Gamma}}}_{ii}$'s are needed for the $N$ elements of $\mathbf{c}$ as seen in (\ref{eq:sbl_weight}).

Recall that the log-sum penalty is equivalent to a Cauchy prior.
So, from the IR-$l_2$ point of view, SBL's formulation in (\ref{eq:sbl_IR}) uses "Cauchy priors" with $N$ parameters ${\tilde{\mathbf{\Gamma}}}_{ii}$'s, while
CG and BLRC's formulation in  (\ref{eq:joint_pdf}) uses Cauchy priors with only one parameter
$\hat{\gamma}$.
In other words, we need to determine $N$ hyper-parameters (i.e., $\tilde{\tau}_i$'s) for characterizing the $N$ parameters ${\tilde{\mathbf{\Gamma}}}_{ii}$'s in SBL, but only one hyper-parameter (i.e., $\hat{\gamma}$) for characterizing the Cauchy prior.
Thus, BLRC leads to a more compact latent space.
This is a major advantage of BLRC over SBL, which can be explained as follows.

Suppose both $c_j$ and $c_l$ in (\ref{eq:linear_model}) are $0$ and to be recovered using SBL and BLRC. During the SBL iteration process, one can see from (\ref{eq:sbl_weight}) that both ${\tilde{\mathbf{\Gamma}}}_{jj}^{(k)}$ and ${\tilde{\mathbf{\Gamma}}}_{ll}^{(k)}$ need to approach $0$ in order to increase the penalties for non-zero $c_j^{(k)}$ and $c_l^{(k)}$ as the iteration index $k$ increases. However, in practice, ${\tilde{\mathbf{\Gamma}}}_{jj}^{(k)}$ and ${\tilde{\mathbf{\Gamma}}}_{ll}^{(k)}$ are  independently controlled by hyper-parameters $\tau_j^{(k)}$ and $\tau_l^{(k)}$. It is very likely that ${\tilde{\mathbf{\Gamma}}}_{jj}^{(k)}$ and ${\tilde{\mathbf{\Gamma}}}_{ll}^{(k)}$ approaches zero with different speeds and therefore the resulting $c_j^{(k)}$ and $c_l^{(k)}$ are not both near $0$. This is not the case with BLRC. In BLRC, as $k$ increases, the only hyper-parameter $\hat{\gamma}^{(k)}$ decreases. It will cause the penalties for non-zero $c_j^{(k)}$ and $c_l^{(k)}$ to increase with more or less the same rate (see (\ref{eq:cg_weight})). Thus, the resulting $c_j^{(k)}$ and $c_l^{(k)}$ will both be near $0$ when the BLRC iteration process converges.

The implication of reducing $\hat{\gamma}^{(k)}$ is shown in the right subplot of Fig.~\ref{fig:GC} 
where $\ln(1+\frac{|c|^2}{\hat{\gamma}^2})$ is plotted for various values of $\hat{\gamma}$.
In addition, the $l_0$ norm, $l_{0.05}$ norm and $l_1$ norm of $c$ are also plotted.
It can be seen that $\ln(1+\frac{|c|^2}{\hat{\gamma}^2})$ approaches $||c||_0$ as $\hat{\gamma}^{(k)}$ approaches zero \cite{shen2013exact}.
This shows that BLRC approximately uses the $l_0$ norm as $\hat{\gamma}^{(k)}$ in the regularization term approaches zero.

\section{Numerical Analyses}
In this section, numerical analyses on numerical stability, convergent properties, spurious targets suppression, noise variance estimation, sensitivities and resolutions
of CG, SBL and BLRC are provided. The OMP \cite{JA07}, which is a popular greedy-based approach for the sparse signal recovery problem, is used as a performance benchmark here.

Consider a sum of $K$ rays for numerical simulation:
\begin{equation}
    y(i)=\sum_{l=1}^{K}  a_l \exp\{j (2 \pi f_l \frac{i}{N}  + \phi_l)\}+\epsilon(i)
     \label{eq:example_1}
\end{equation}
where $i\in S$ and $S \subset \{1,2,...,N \}$. The number of elements in $S$ is $M$ which is smaller than $N$. Here. $\{f_l\}$, {$\{a_l\}$} and {$\{\phi_l\}$}  are spatial frequencies, amplitudes, and phases, respectively.
The additive noise {$\{\epsilon(i)\}$} are independent identical distributed zero-mean Gaussian random variables with variance $\sigma_n^2$. 

\subsection{A Six-Ray Example}

\begin{figure}[!t]
\centering
  \includegraphics[width=3.3in]{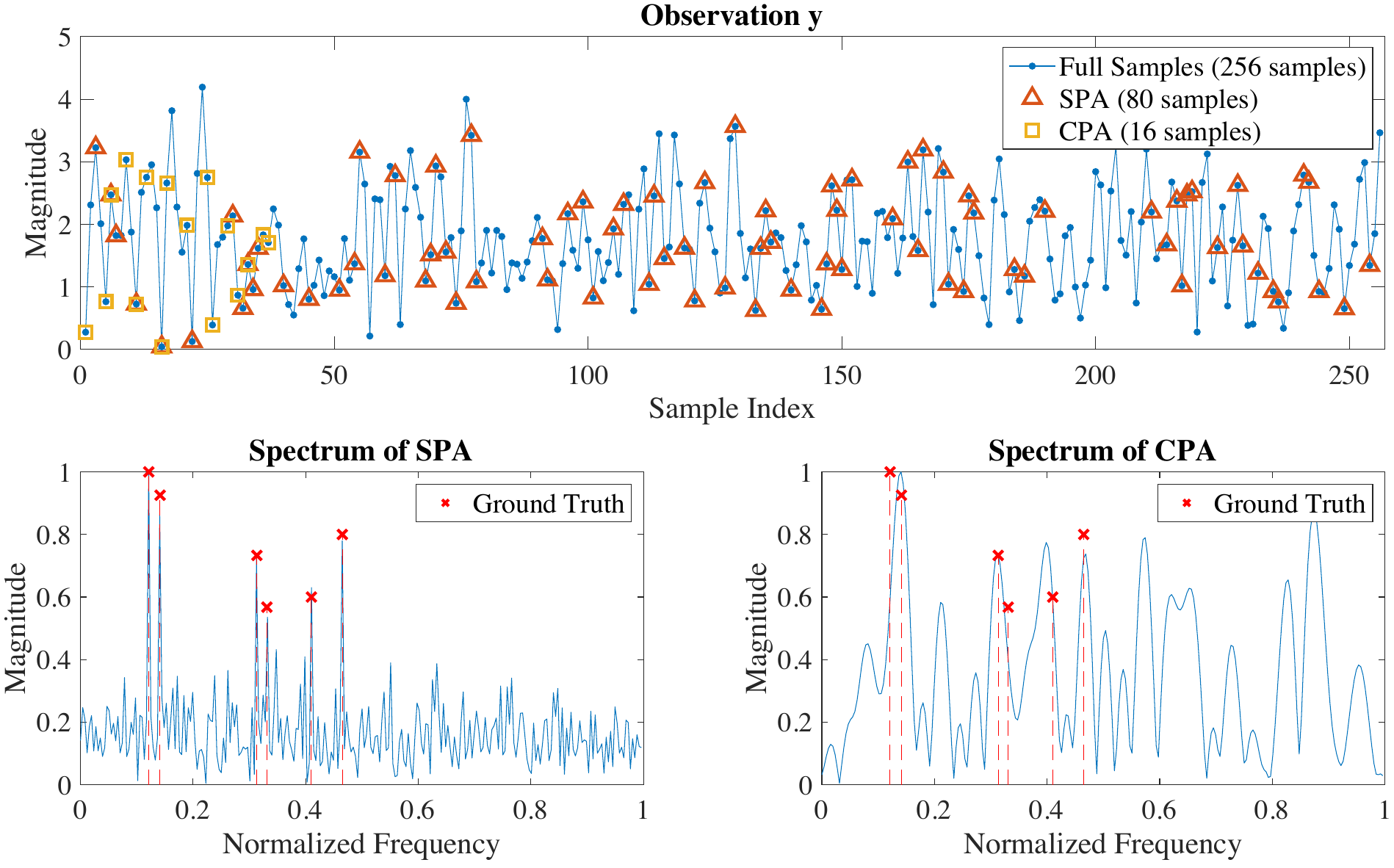}
  \caption{Time samples and Spectra of sparse array (SPA) and coprime array (CPA)}
  \label{fig:dft_omp3}
\end{figure}

Without loss of generality, we choose $N=256$.
The arbitrarily selected spatial frequencies {$f_l$}, amplitudes {$a_l$}, and phases {$\phi_l$} for $K=6$ are listed in Table I. The standard deviation of the additive noise {$\epsilon(i)$} is ${\sigma}_n = 0.1$.
Consider a large sparse antenna array (SPA) with $M=80$ and a small coprime array (CPA) with $M=16$.
The 80 random samples observed by the large SPA, the 16 samples observed by the small CPA, and 
the 256 full samples of $\{|y(i)|\}$ are shown in the top subplot of Fig. \ref{fig:dft_omp3}.
Fourier spectra of SPA and CPA are shown in the bottom left and right subplots, respectively, of Fig. \ref{fig:dft_omp3}
where the six rays are marked as ground truth.
It is seen that the six rays are fully resolved by the large SPA. However, many small noise-like side lobes are generated.
For the small CPA, two rays are missing (unresolved). In addition, several large spurious rays appeared.

\begin{table}[!t]
\renewcommand{\arraystretch}{1.3}
\centering
\caption{\textsc{simulation example: six rays}}
\label{tab:y}
\begin{tabular}{|c||c|}
%\midrule
\hline
ray indexes  & $[1,~~2,~~3,~~4,~~5,~~6]$ \\
\hline
Frequencies & $[0.1212,~ 0.1413,~0.3132,~0.331,~0.41,~0.465]$ \\
\hline
Amplitudes	& $[1,~0.9254,~0.7331,~0.5678,~0.6,~0.8]$ \\
\hline
Phases & $[5.1191,~5.6913,~0.7979,~5.7389,~3.9732,~0.6129]$ \\
\hline
\end{tabular}
\end{table}

Using the large SPA and small CPA shown in Fig. \ref{fig:dft_omp3}, OMP, CG, SBL, and BLRC are employed to recover the sparse signal defined in (\ref{eq:example_1}) and Table I.
Note that the Bayesian approaches (CG, SBL, and BLRC) summarized in Section II have been formulated in terms of real variables and parameters.
The extension of these approaches to deal with complex variables and parameters is straightforward as shown in \cite{DP07}. 
At first, the $(\cdot)^2$ in Cauchy distribution (\ref{eq:cauchy_prior}) is replaced by $|(\cdot)|^2$, and the likelihood distribution in (\ref{eq:likelihood}) is changed to complex Gaussian.
Then, the derivations are the same as before except that we need to change transpose $(\cdot)^T$ to Hermitian transpose $(\cdot)^H$ and replace $(\cdot)^2$ by $|(\cdot)|^2$. 

\subsection{Initial Values of Hyper-parameters}
SBL is robust with respect to the selection of initial values. Here, the initial values are $\tilde{\sigma}_n = 1$ and $\tilde{\sigma}_i=0.1,~\forall i$ for both SPA and CPA. 

For CG, the selection of the optimal values of $\hat{\sigma}_n$ and $\hat{\gamma}$ is crucial to the performance of CG. 
Based on the knowledge of the targets in Table I, the optimal values chosen by trial and error are $\hat{\sigma}_n = 0.1$ and $\hat{\gamma}=0.01$ for both SPA and CPA in our simulation. 
However, we are not able to know the true targets in practice. 
Therefore, it is impossible to find the optimal values for $\hat{\sigma}_n$ and $\hat{\gamma}$ for CG in reality.
The initial estimate of each element of $\mathbf{c}$ for  CG is a random number uniformly distributed in $[0~1]$. 

Like SBL, the selection of initial $\hat{\sigma}_n$ and $\hat{\gamma}$ for BLRC is not critical since they will be updated robustly as the iteration proceeds.
The rule of thumb is that the initial $\hat{\gamma}$ needs to be large enough to emphasize the data fitting term $||\mathbf{y-Ac}||^2$ in (\ref{eq:joint_pdf}) at the beginning of the iteration process. 
As the iteration proceeds, $\hat{\gamma}$ will continue to decrease so as to emphasize the regularization term $ {\hat{\sigma}_n^2}\sum_{i=1}^{N}2\ln({c_i^2}+{\hat{\gamma}^2})$ in (\ref{eq:joint_pdf}) in order to promote sparsity.
However, $\hat{\gamma}$ should not be too large, otherwise it will require more iterations to converge.
Since there are fewer measurements with CPA ($M=16$) compared with SPA ($M=80$), there exist more local minimums in CPA \cite{Wipfphd}.
Thus, we need to have a larger initial value of $\hat{\gamma}$ with CPA to avoid being trapped in a local minimum in the early stage of iterations. In our simulations, the initial values for BLRC are $\hat{\sigma}_n = 1$ and $\hat{\gamma}=0.1$ for SPA and 
$\hat{\sigma}_n = 1$ and $\hat{\gamma}=1$ for CPA. 
Like CG, the initial estimate of each element of $\mathbf{c}$ for BLRC is also a random number uniformly distributed in $[0~1]$.

\begin{figure}[!t]
\centering
  \includegraphics[width=3.3in]{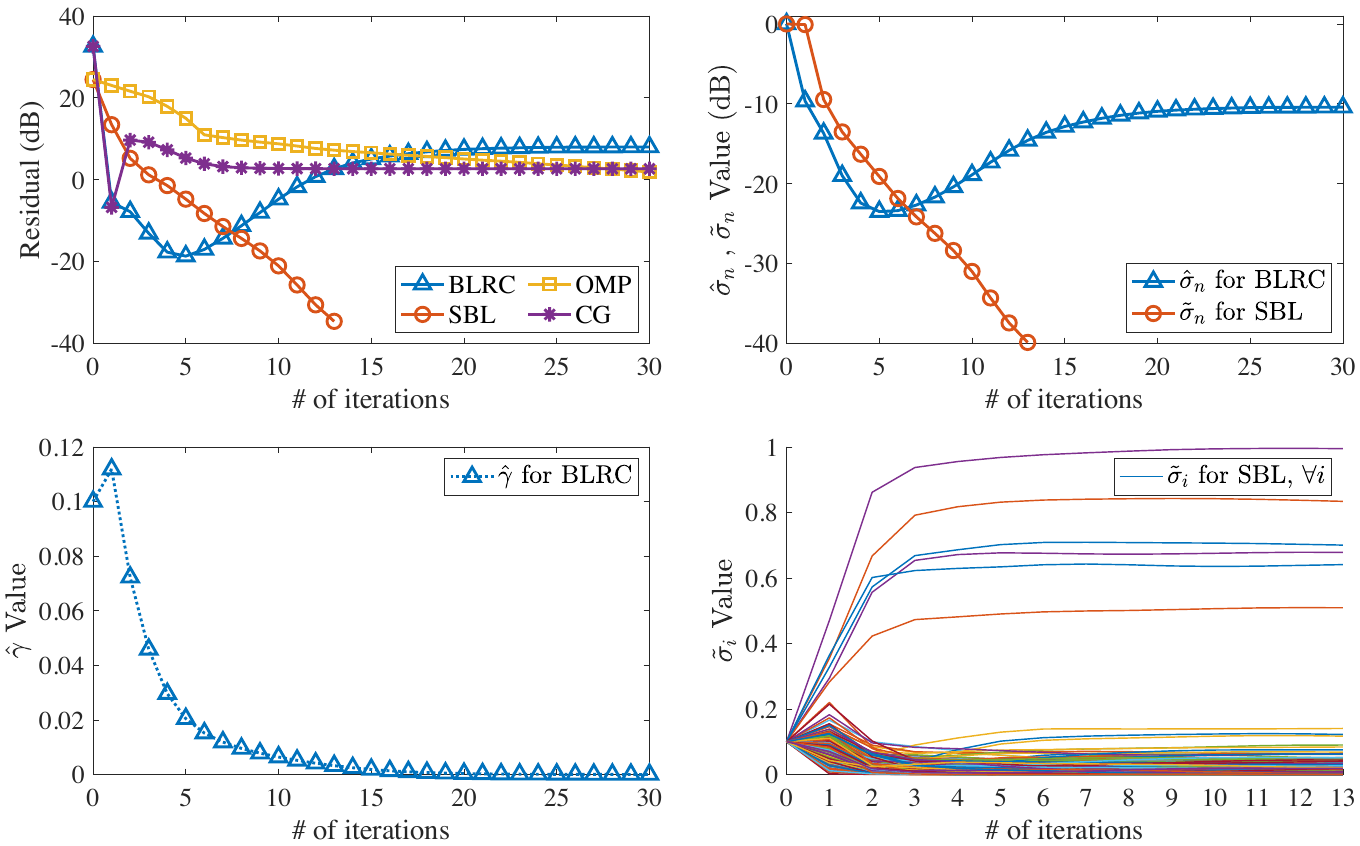}
  \caption{Convergent Properties of OMP, SBL, CG, and BLRC using SPA}
  \label{fig:NR1_2_2}
\end{figure}

\begin{figure}[!t]
\centering
  \includegraphics[width=3.3in]{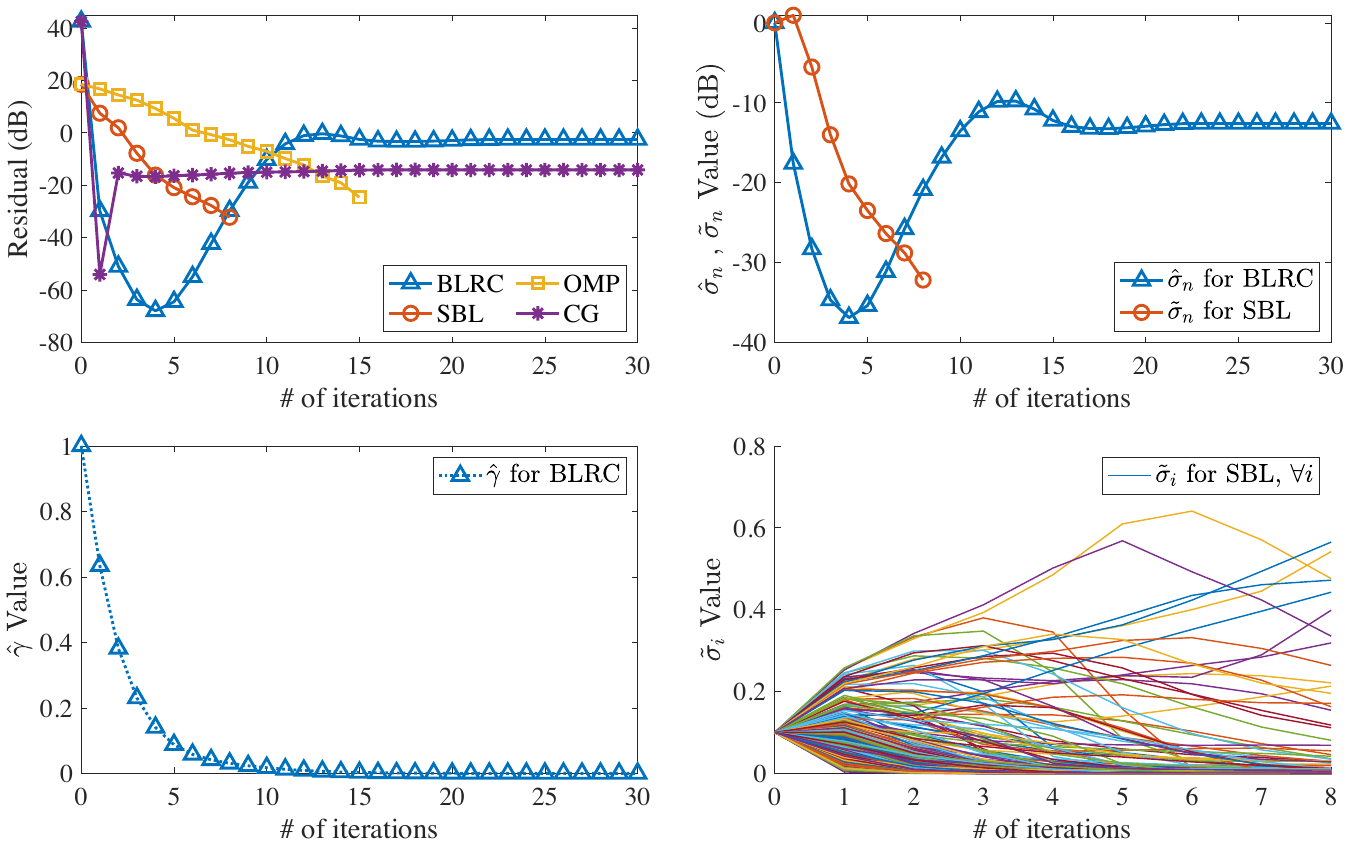}
  \caption{Convergent Properties of OMP, SBL, CG, and BLRC using CPA}
  \label{fig:NR3_CPA_convergent}
\end{figure}

Using SPA and CPA, respectively, Fig.~\ref{fig:NR1_2_2} and Fig.~\ref{fig:NR3_CPA_convergent} show the values of key parameters versus the iteration number for OMP, CG, SBL, and BLRC.
These results will be used to discuss various numerical aspects of the four considered approaches in the following subsections.
Here, in the top left subplots of both figures, the residue in dB is defined as $R_{dB}^{(k)} \equiv 20 \log_{10}||\mathbf{y-Ac}^{(k)}||$ where $\mathbf{c}^{(k)}$ is the estimate of $\mathbf{c}$ using the corresponding approach in the $k^{th}$ iteration.

\begin{figure}[!t]
\centering
  \includegraphics[width=3.3in]{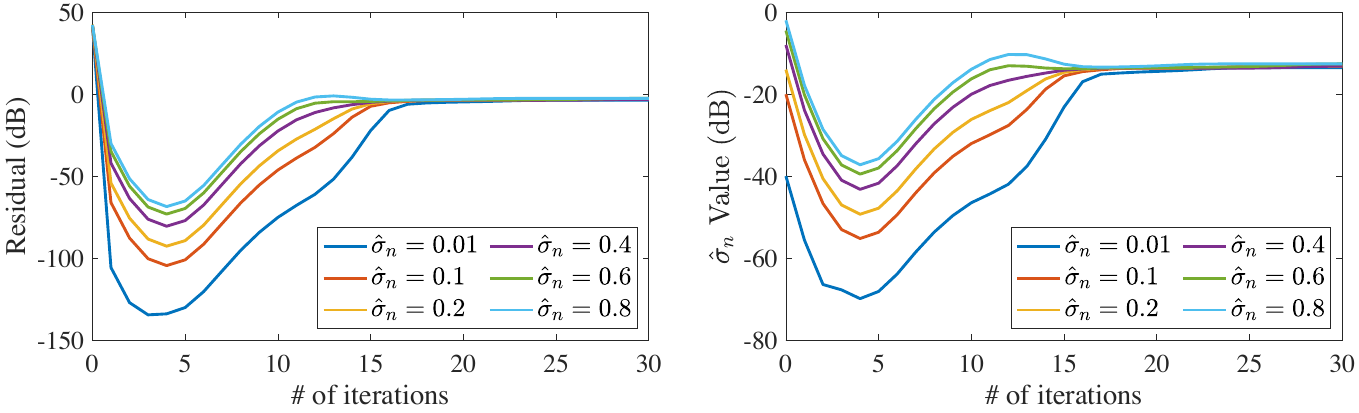}
  \caption{Convergent Properties of BLRC for CPA with different initial values of $\hat{\sigma}_n$}
  \label{fig:ini_sigma_n}
\end{figure}

To further demonstrate that BLRC is robust to the initial values of hyper-parameters,
numerical simulations with different initial values of $\hat{\sigma}_n$ and $\hat{\gamma}$ are performed for CPA.
In the left subplot of Fig.~\ref{fig:ini_sigma_n}, the residues for BLRC with different initial $\hat{\sigma}_n$ values are plotted against the iteration number while the initial value of $\hat{\gamma}$ is $1$.
For these initial values, the updated $\hat{\sigma}_n$ values are plotted against the iteration number in the right subplot of Fig.~\ref{fig:ini_sigma_n}.
Although different initial values are set for $\hat{\sigma}_n$, identical final residue, final estimate of $\hat{\sigma}_n$, and the recovery result for $\mathbf{c}$ are obtained. 

\begin{figure}[!t]
\centering
  \includegraphics[width=3.3in]{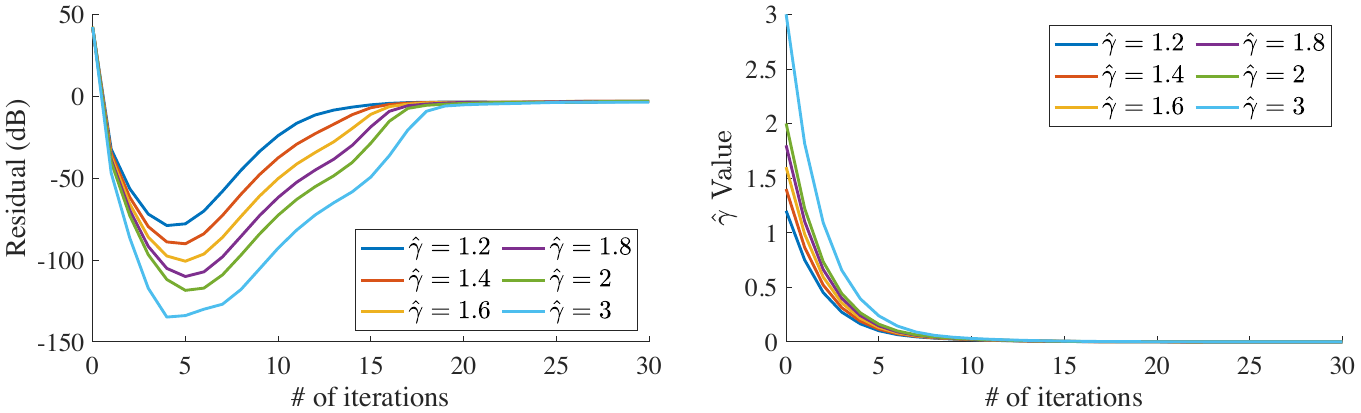}
  \caption{Convergent Properties of BLRC for CPA with different initial values of $\hat{\gamma}$}
  \label{fig:ini_gamma}
\end{figure}

In the left subplot of Fig.~\ref{fig:ini_gamma}, the residues for BLRC with different initial $\hat{\gamma}$ values are plotted against the iteration number while the initial value of $\hat{\sigma}_n$ is $1$.
For these initial values, the updated $\hat{\gamma}$ values are plotted against the iteration number in the right subplot of Fig.~\ref{fig:ini_gamma}.
Similarly, although we have different initial values for $\hat{\gamma}$, identical final residue, final estimate of $\hat{\gamma}$, and recovery result for $\mathbf{c}$ are obtained.

\subsection{Numerical Stability}

In Fig.~\ref{fig:NR1_2_2} and Fig.~\ref{fig:NR3_CPA_convergent}, the iterative procedure of SBL is terminated at $k=13$ and $k=8$ for SPA and CPA, respectively.
This is due to the fact that the matrix  $\mathbf{H}_{sbl}^{(k)} = \frac{1}{(\tilde{\sigma}_n^{(k-1)})^2}
 \mathbf{A}^T\mathbf{A}+\tilde{\mathbf{\Sigma}}^{(k)} $ becomes ill-conditioned for large $k$'s.
Note that $\mathbf{H}_{sbl}^{(k)}$ needs to be inverted so as to obtain the posterior covariance matrix $\tilde{\mathbf{\Gamma}}^{(k)}$ in step 4 of Algorithm 2.
Unlike SBL,
the matrix $ \mathbf{H}_{blrc}^{(k)} $, defined as $ \frac{1}{(\hat{\sigma}_n^2)^{(k-1)}}\mathbf{A}^T\mathbf{A}
    +\frac{2}{(\hat{\gamma}^{(k-1)})^2}\hat{\mathbf{Q}}^{(k)} $, does not become ill-conditioned for the same set of $k$'s.
Note that $\mathbf{H}_{blrc}^{(k)}$  is to be inverted for obtaining the posterior covariance matrix $\hat{\mathbf{\Gamma}}^{(k)}$ in step 4 of Algorithm 3.
The condition numbers of $\mathbf{H}_{sbl}^{(k)}$ and $\mathbf{H}_{blrc}^{(k)}$ for some $k$'s are also shown in Table~\ref{tab:SBL} and Table~\ref{tab:cond_CPA} for SPA and CPA, respectively.

\begin{table}[!t]
\renewcommand{\arraystretch}{1.3}
\centering
\caption{\textsc{Condition Numbers with SPA - SBL and BLRC}}
\label{tab:SBL}
\begin{tabular}{|c||c|c|c|c|c|}
%\midrule
\hline
$k$  & $5$ & $13$ & $14$ & $25$ \\
\hline
$\mathbf{H}_{sbl}^{(k)}$ & $3.4\times 10^6$ & $8.1\times 10^{14}$ & \rm{N/A} & \rm{N/A}  \\
\hline
$\mathbf{H}_{blrc}^{(k)}$ & $133$ & $130$ & $236$ & $8.7\times 10^5$\\
\hline
\end{tabular}
\end{table}

\begin{table}[!t]
\renewcommand{\arraystretch}{1.3}
\centering
\caption{\textsc{Condition Numbers with CPA - SBL and BLRC}}
\label{tab:cond_CPA}
\begin{tabular}{|c||c|c|c|c|}
%\midrule
\hline
$k$  & $5$ & $8$ & $9$ & $25$ \\
\hline
$\mathbf{H}_{sbl}^{(k)}$ & $5.4\times 10^{9}$ & $2.9\times 10^{13}$ & \rm{N/A} & \rm{N/A} \\
\hline
$\mathbf{H}_{blrc}^{(k)}$ & $1.6\times 10^4$ & $1.1\times 10^3$ & $734$ & $8.5\times 10^6$  \\
\hline
\end{tabular}
\end{table}

\begin{figure}[!t]
\centering
  \includegraphics[width=3.3in]{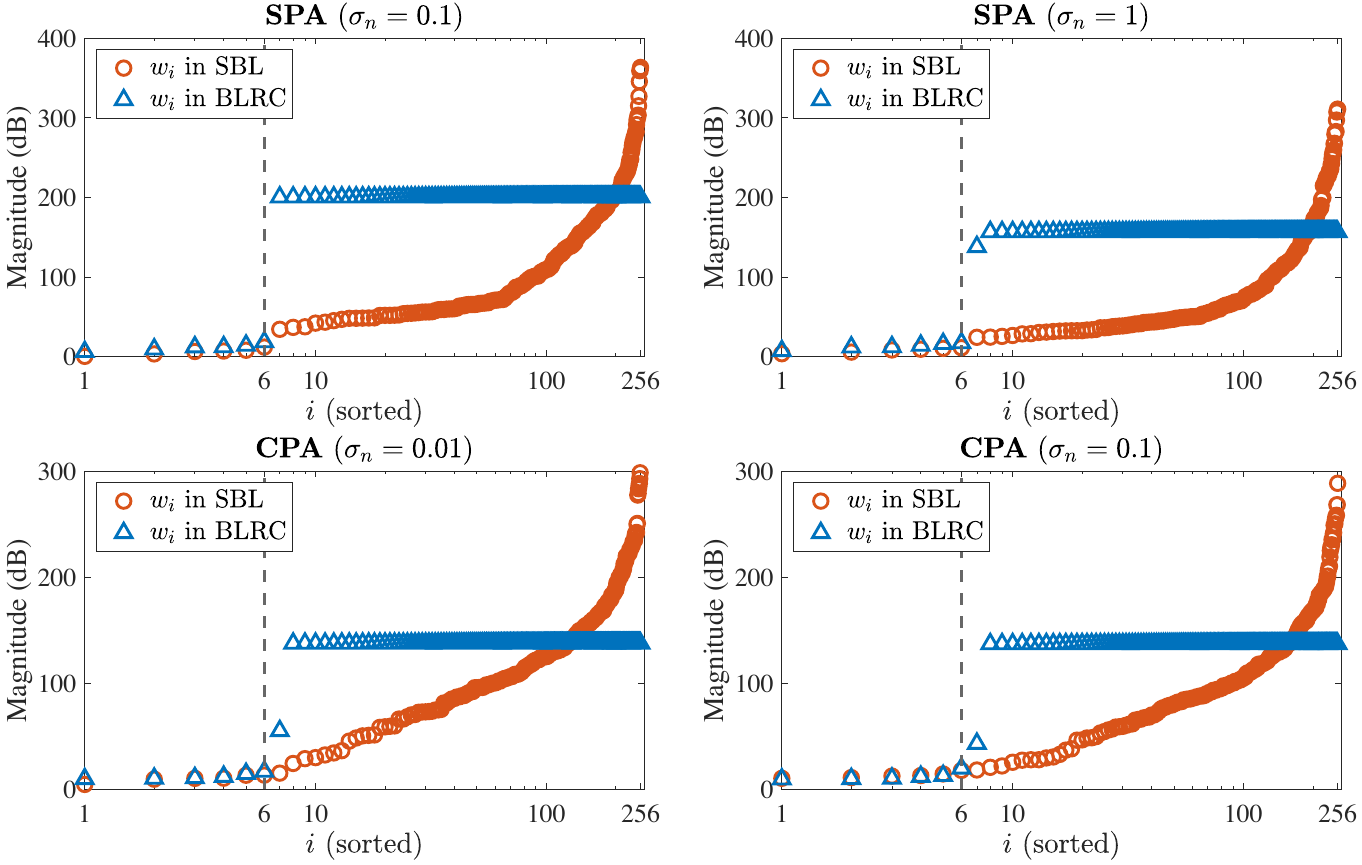}
  \caption{Sorted magnitudes in dB of the IR-$l_2$ weights, $20 \text{log}_{10} |w_i^{(k)}|$, derived from (\ref{eq:cg_weight}) for BLRC and (\ref{eq:sbl_weight}) for SBL, $\forall i=1,...,N$. The dashed vertical line at $i=6$ marks the number of true targets}
  \label{fig:weightc}
\end{figure}

To understand the reasons why BLRC is numerically more stable than SBL, magnitudes of the IR-$l_2$ weights $\{w_i^{(k)}\}$ in (\ref{eq:cg_weight})
and (\ref{eq:sbl_weight}) for BLRC and SBL, respectively, are sorted in ascending order (see Fig.~\ref{fig:weightc}).
Note that in Fig.~\ref{fig:weightc} the $y$-axis is represented in dB so as to emphasize the small $|w_i^{(k)}|$ values.
The $x$-axis is also presented in log scale to emphasize the small $i$'s.

Recall that the IR-$l_2$ weights $\{w_i^{(k)}\}$ in (\ref{eq:cg_weight}) are the diagonal elements of  $2\hat{\mathbf{Q}}^{(k)}/(\hat{\gamma}^{(k-1)})^2$ in step 4 of Algorithm 3. And the weights $\{w_i^{(k)}\}$ in (\ref{eq:sbl_weight}) are equal to the diagonal elements of $\tilde{\mathbf{\Sigma}}^{(k)}$ in step 4 of Algorithm 2.
Since $\mathbf{A^TA}$ is singular, a well-conditioned diagonal loading of $\tilde{\mathbf{\Sigma}}^{(k)}$ is desirable so as to make $\mathbf{H}_{sbl}^{(k)}$ non-singular.
Similarly, a well-conditioned diagonal loading of $2\hat{\mathbf{Q}}^{(k)}/(\hat{\gamma}^{(k-1)})^2$ is desirable in order to make $\mathbf{H}_{blrc}^{(k)}$ invertible.

From Fig.~\ref{fig:weightc},
one can see that the ratio between maximum $|w_i^{(k)}|$ and minimum $|w_i^{(k)}|$ is near or over $10^{15}$ for SBL. Moreover, $|w_i^{(k)}|$ is more or less uniformly distributed between maximum $|w_i^{(k)}|$ and minimum $|w_i^{(k)}|$ for SBL.
This is due to the fact that each element of $\{c_i\}$ is characterized by one hyper-parameter $\tilde{\sigma_i}$ in SBL. When a particular $c_i$ is near zero, the variance $\sigma_i^2$(i.e., $\frac{1}{\tau_i}$ or $\frac{1}{w_i}$) of the corresponding zero-mean Gaussian distribution must approach zero. Thus, $\tau_i$ can be seen as the confidence of “the corresponding $c_i$ is zero”. For different elements of $\{c_i\}$, we typically have different confidence levels on whether they are zero or not. In addition, the confidence levels change as iteration increases. Thus, the elements of $\tau$ (i.e., the $w_i$) approach infinity at various speeds during iterations and the diagonal loading for $\mathbf{H}_{sbl}^{(k)}$ (i.e., $\tilde{\mathbf{\Sigma}}^{(k)}$ ) is near singular. As a consequence, $\mathbf{H}_{sbl}^{(k)}$ is also near singular as shown in Table~\ref{tab:SBL} and Table~\ref{tab:cond_CPA}, which is also noticed by the SBL original paper \cite{sblrvm}. Although we could use $\sigma_i^2$ instead of the precision parameters (i.e., $\tau_i$) as in \cite{liao2022map}, this would not resolve the singularity issue of the matrix $\mathbf{H}_{sbl}^{(k)}$ here. This is because $\sigma_i^2$ would still be inverted in the subsequent step of SBL algorithm, leaving the core problem unaddressed.

Unlike SBL, 
the ratio between maximum $|w_i^{(k)}|$ and minimum $|w_i^{(k)}|$ is near or less than $10^{10}$ for BLRC. Most importantly, $|w_i^{(k)}|$ is more or less constant for BLRC. This is due to the fact that only one scale parameter $\hat{\gamma}$ is used for characterizing  $\{c_i\}$.
Since most $\{c_i\}$ are near zeros when BLRC converges, most $\{w_i^{(k)}\}$ are determined only by $\hat{\gamma}$ as shown in  (\ref{eq:cg_weight}) and are therefore almost constant.
Now that, as the diagonal loading for $\mathbf{H}_{blrc}^{(k)}$ (i.e., $2\hat{\mathbf{Q}}^{(k)}/(\hat{\gamma}^{(k-1)})^2$) is almost constant,
$\mathbf{H}_{blrc}^{(k)}$ is not near singular as shown in Table~\ref{tab:SBL} and Table~\ref{tab:cond_CPA}. This demonstrates that BLRC is more numerically robust and stable than SBL with sparse array input, making it more suitable for automotive radar applications, which commonly utilize sparse arrays and operate under single-precision computing environments. Moreover, the almost constant diagonal property of  $2\hat{\mathbf{Q}}^{(k)}/(\hat{\gamma}^{(k-1)})^2$ helps in noise variance estimation (which will be discussed in Section VI.F).

\subsection{Residues and Convergent Criteria}

In the top left subplots in Fig.~\ref{fig:NR1_2_2} for SPA and
Fig.~\ref{fig:NR3_CPA_convergent} for CPA, we show the residues in dB, $R_{dB}^{(k)}$, derived from the $k^{th}$ iteration.

The residue for OMP decreases as the iteration number $k$ increases. 
However, the decreasing rate reduces as $k>K=6$ since all six rays have been identified at $k=K$. 
The new rays estimated by OMP after $k=K$ are false targets derived from matching the noises.
Since the further decrease of residue does not promote sparsity, 
we should stop the OMP iteration when the residue is less than an appropriate threshold $t_r$
for practical implementations.
Without knowing the number of targets in advance, the OMP estimation will most likely either miss true targets or show false targets. 
Note that the OMP iteration stops at the $15^{th}$ iteration in Fig.~\ref{fig:NR3_CPA_convergent} since the number of measurements $M=16$ for CPA in our example.

The residue for SBL decreases rapidly as the iteration number $k$ increases. 
Although the residue is very small for large $k$'s, it does not mean the estimates of $\mathbf{c}$ obtained for 
large $k$'s are better than those obtained for smaller $k$'s. 
This is because 
the posterior covariance matrix $\tilde{\mathbf{\Gamma}}$ in step 4 of Algorithm 2
becomes ill-conditioned when $k>13$ for SPA 
(see Table~\ref{tab:SBL}) and when $k>8$ for CPA (see Table~\ref{tab:cond_CPA}).
Moreover, further decrease of the residue 
only adjusts the estimated $\mathbf{c}$ to minimize the mismatch between $\mathbf{y}$ and $\mathbf{Ac}$, which is due to the additive noises.
It does not promote sparsity. This phenomenon is consistent with the findings in \cite{Wipf11} which shows that SBL will converge to a $l_0$ regularized optimization problem when there is no noise.
Thus, for practical implementations,
we should stop the SBL iteration when the residue is less than an appropriate threshold $t_r$.
In addition, the SBL iteration stops when
the posterior covariance matrix $\tilde{\mathbf{\Gamma}}$ in step 4 of Algorithm 2
becomes ill-conditioned  (see
Table~\ref{tab:SBL} for SPA and Table~\ref{tab:cond_CPA} for CPA).

Similar to CG, the residue for BLRC does not always decrease as the iteration number $k$ increases.
At the beginning of iterations, like SBL, the residue for BLRC  decreases as the iteration number $k$ increases. This is because $\hat{\gamma}$ is large and the data fitting term $||\mathbf{y-Ac}||^2$ in (\ref{eq:joint_pdf}) is emphasized in this phase.
As the iteration proceeds, $\hat{\gamma}$ continues to decreases (see the bottom left subplots in Fig.~\ref{fig:NR1_2_2} for SPA and Fig.~\ref{fig:NR3_CPA_convergent} for CPA).
Then,  the regularization term $ {\hat{\sigma}_n^2}\sum_{i=1}^{N}2\ln({c_i^2}+{\hat{\gamma}^2})$ in (\ref{eq:joint_pdf}) becomes more and more important in order to promote sparsity.
Although the residue could increase in this phase, the rate of residue change commonly decreases as the $k$ increases and
the residue gradually becomes nearly a constant.
This shows that BLRC not only minimizes the residue but also promotes sparsity.
Thus, like CG, we should stop the BLRC iteration if the absolute value of the residue change is less than an appropriate threshold for practical implementations.
Note that, unlike SBL,
the posterior covariance matrix $\hat{\mathbf{\Gamma}}$ in step 4 of Algorithm 3
does not become ill-conditioned for relatively large $k$'s (see
Table~\ref{tab:SBL} for SPA and Table~\ref{tab:cond_CPA} for CPA).

\subsection{Sparse Spectrum Reconstruction and Side Lobes}

Note that the SPA with $M=80$ has a much larger aperture than the CPA with $M=16$.
As it has been shown in \cite{Wipfphd} that
larger $\frac{N}{M}$ ratios imply more local minimums,
it will be much easier to reconstruct the original spectrum using SPA than CPA.  
The CPA example puts the four approaches under a severe test.

\subsubsection{Sparse Array (SPA) with $M=80$}

\begin{figure}[!t]
\centering
  \includegraphics[width=3.3in]{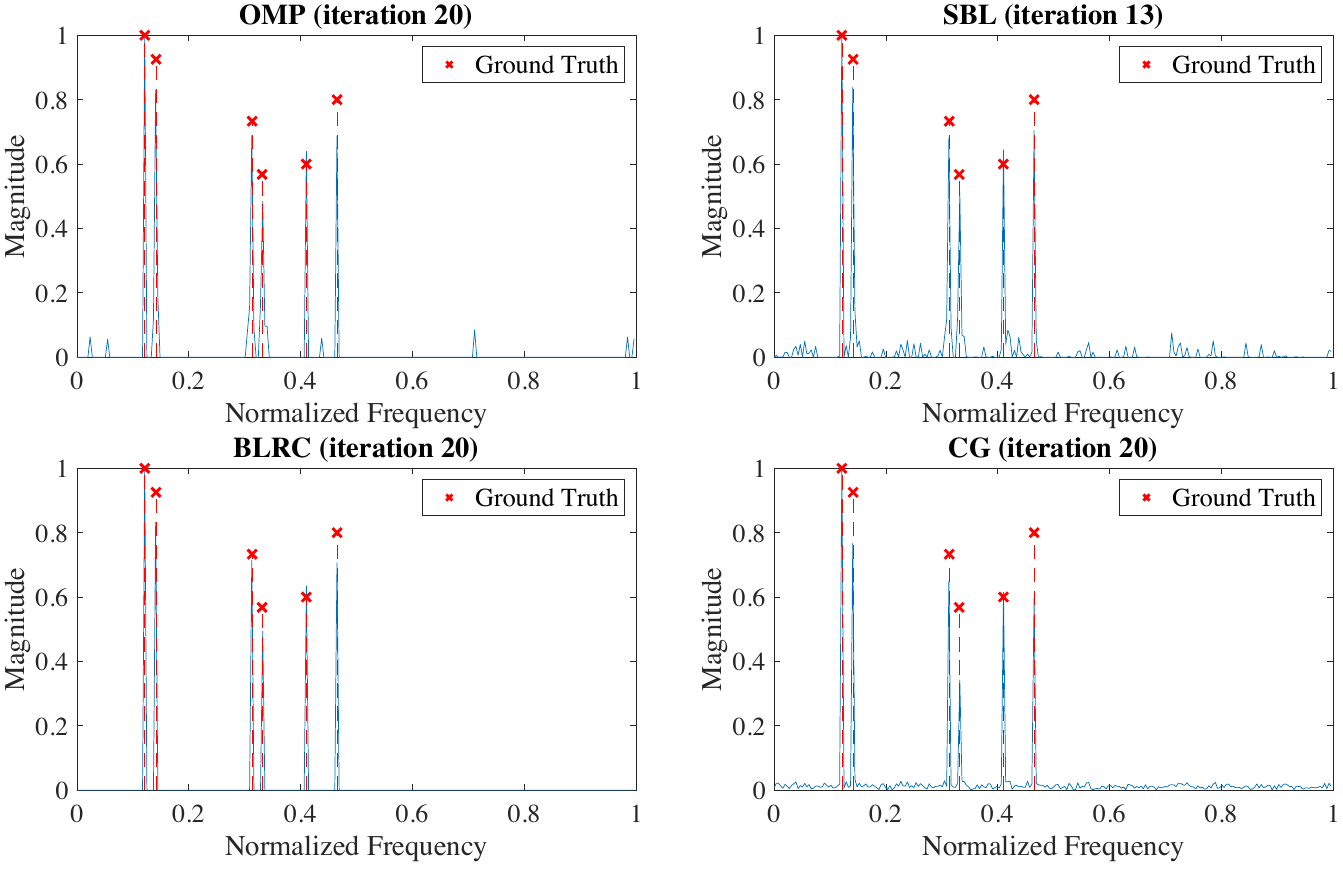}
  \caption{Recovery of $\mathbf{c}$ at iteration 20 using SPA (Note that the SBL stops at iteration 13) with $\sigma_n=0.1$}
  \label{fig:stopend}
\end{figure}

\begin{figure}[!t]
\centering
  \includegraphics[width=3.3in]{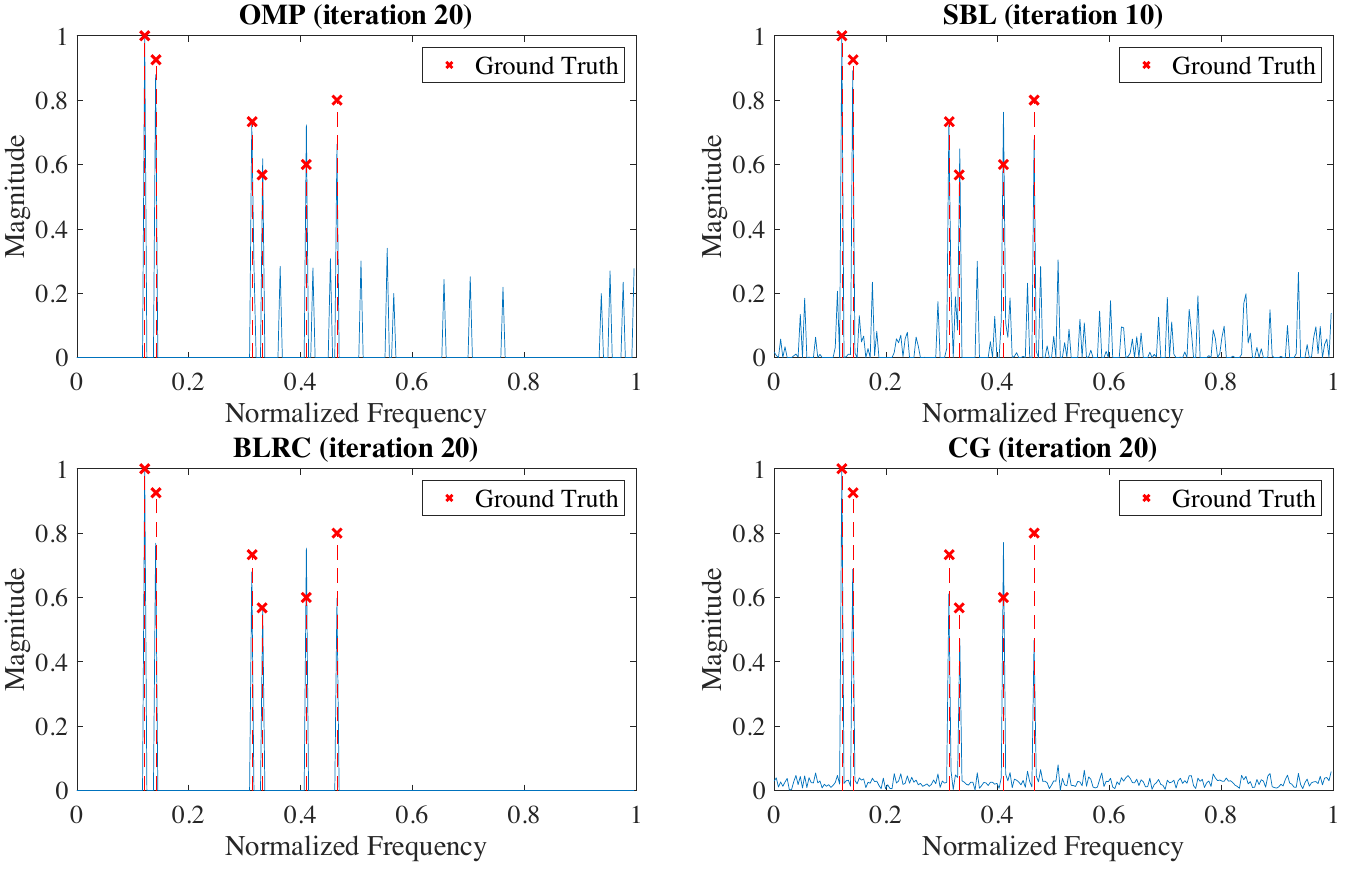}
  \caption{Recovery of $\mathbf{c}$ at iteration 20 using SPA (Note that the SBL stops at iteration 10) with $\sigma_n=1$}
  \label{fig:stop1}
\end{figure}

Using the SPA samples in Fig. \ref{fig:dft_omp3}, Fig.~\ref{fig:stopend} and Fig.~\ref{fig:stop1} show the estimated $\mathbf{c}$ with $\sigma_n=0.1$ and $\sigma_n=1$, respectively.

As shown in Fig.~\ref{fig:stopend}, OMP, SBL, BLRC, and CG all recover six rays with high frequency accuracy.
The magnitudes of the six rays recovered by CG are the most inaccurate among the four approaches. 
Regarding side lobes, OMP suggests twenty rays at $k=20$ and therefore generates $14$ distinct side lobes.
Both SBL and CG have many small side lobes where
the side lobes for CG are like white noises and the side lobes for SBL are like weak rays.
It is remarkable that most side lobes are suppressed by BLRC.

The sparse signal recovery properties of SBL and BLRC with SPA can be explained using the two bottom subplots in  Fig.~\ref{fig:NR1_2_2}.
For SBL in the bottom right subplot, the six top curves with  $\tilde{\sigma}_i>0.4$ (after the $4^{th}$ iteration) represent the six true targets.
The rest of the curves with small but non-negligible $\tilde{\sigma}_i$'s represent the false targets.
It can be seen that SBL has many spurious targets.
For BLRC in the bottom left subplot, $\hat{\gamma}$ decreases drastically to promote sparsity as $k$ increases.
This is due to the fact that a smaller $\hat{\gamma}^{(k)}$ makes the regularization term in (\ref{eq:joint_pdf}) closer to the $l_0$ norm, as shown in the right subplot of Fig.~\ref{fig:GC}.

The sparse signal recovery properties of SBL and BLRC with SPA can also be seen from the two top subplots in Fig.~\ref{fig:weightc}.
For both SBL and BLRC, the leftmost six points ($i=1,2...,6$) with the six smallest weights represent the six true targets.
Those points in the middle and the right side of the two top subplots are with very large weights and will generate essentially zero $c_i$'s.
The rest of the points with moderate weights (which are not large enough to generate negligible $c_i$'s) will be considered as false targets.
It can be seen that SBL has many more spurious targets than BLRC because many weights of SBL are not large enough to generate negligible $c_i$'s. 

When $\sigma_n$ increases from $0.1$ to $1$, the magnitudes of false targets derived from OMP, SBL, and CG increase drastically (comparing Fig.~\ref{fig:stopend} with Fig.~\ref{fig:stop1}).
However, the spurious targets are suppressed by BLRC for both $\sigma_n=0.1$ and $\sigma_n=1$. Since the largest ray magnitude is 1 as shown in Table I, it is concluded that BLRC can suppress spurious targets in low signal-to-noise ratio (SNR).

\subsubsection{Coprime Arrays (CPA) with $M=16$}

\begin{figure}[!t]
\centering
  \includegraphics[width=3.3in]{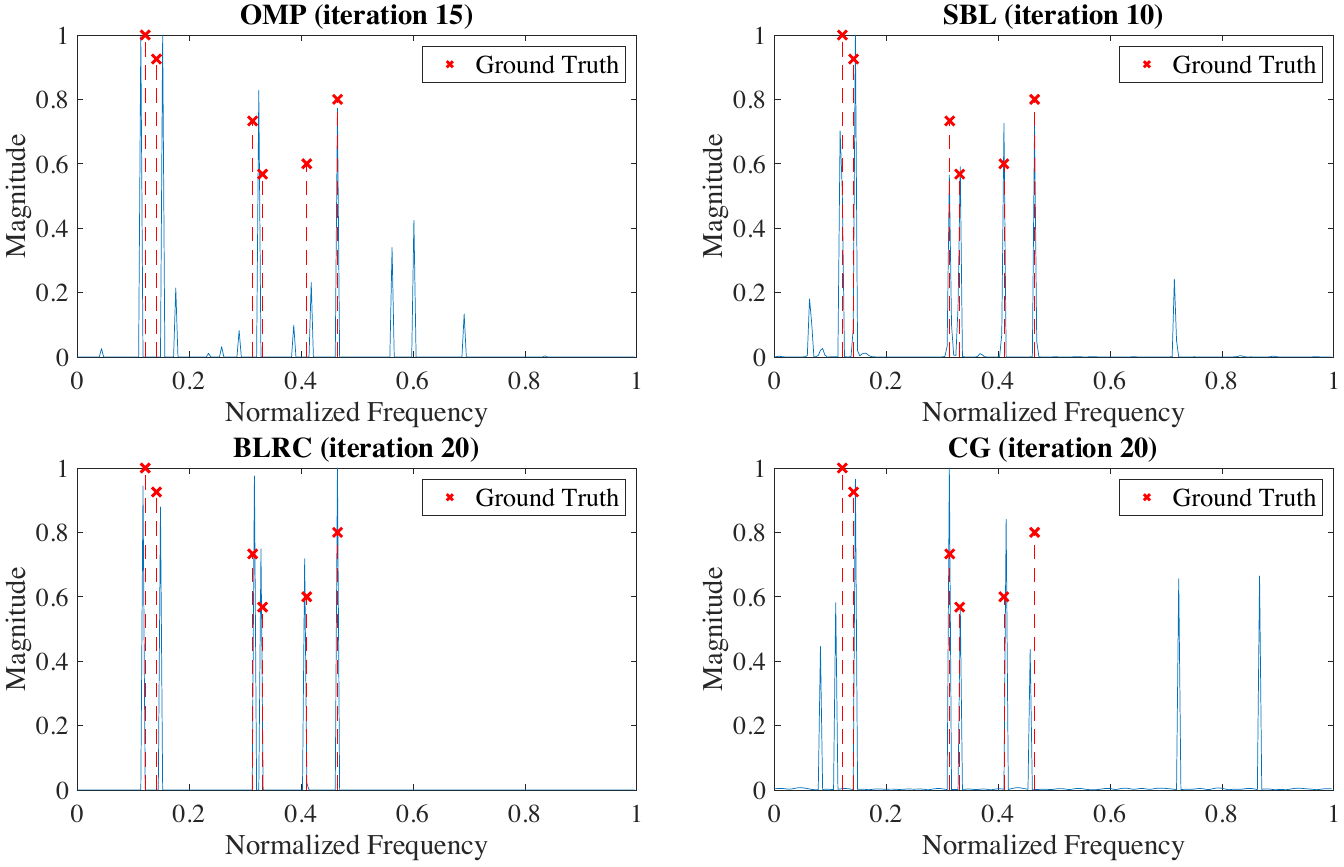}
  \caption{Recovery of $\mathbf{c}$ at iteration 20 using CPA (Note that the OMP stops at iteration 15 and SBL stops at iteration 10) with $\sigma_n=0.01$}
  \label{fig:coprime1}
\end{figure}

\begin{figure}[!t]
\centering
  \includegraphics[width=3.3in]{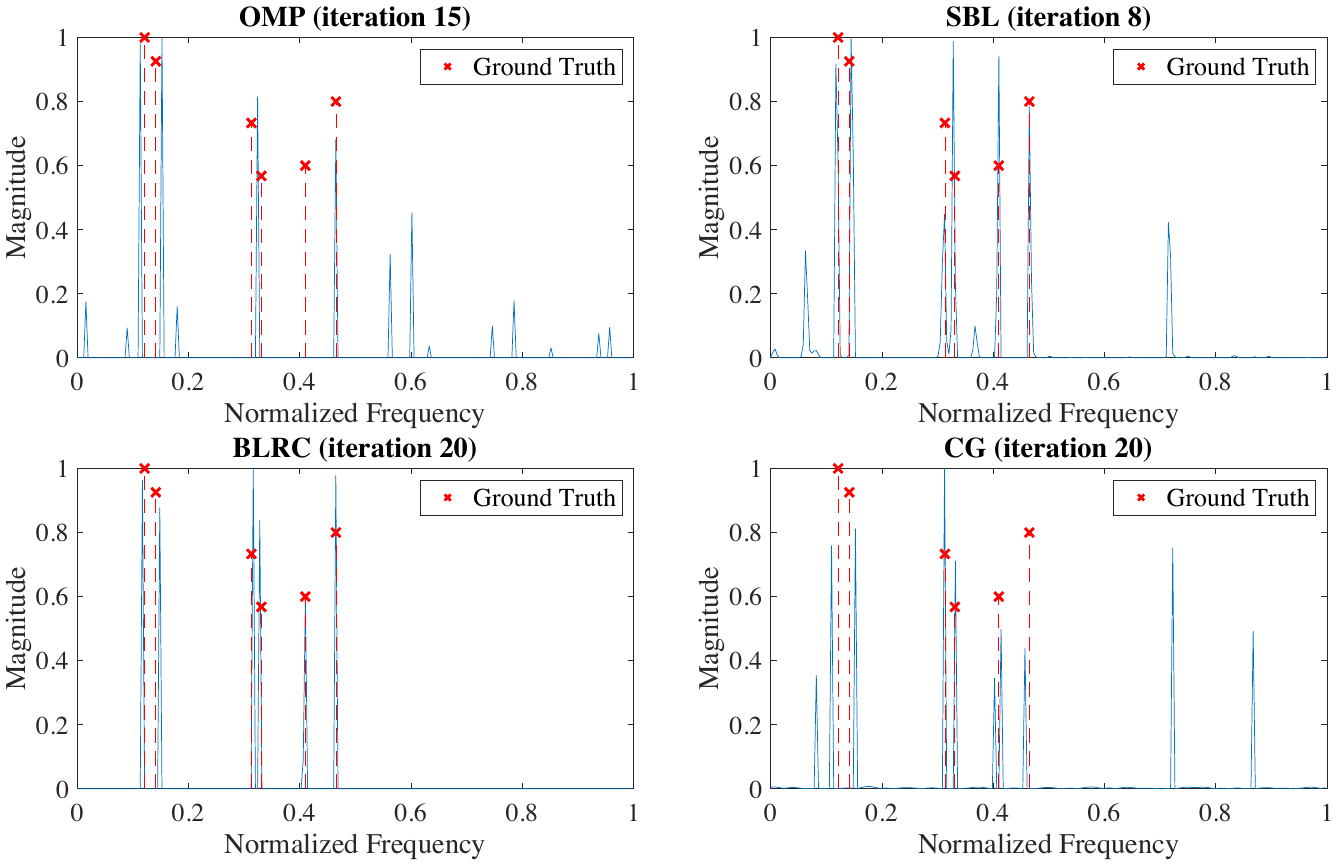}
  \caption{Recovery of $\mathbf{c}$ at iteration 20 using CPA (Note that the OMP stops at iteration 15 and SBL stops at iteration 8) with $\sigma_n=0.1$}
  \label{fig:coprime2}
\end{figure}

Using the CPA samples in Fig. \ref{fig:dft_omp3}, Fig.~\ref{fig:coprime1} and Fig.~\ref{fig:coprime2} show the estimated $\mathbf{c}$ with $\sigma_n=0.01$ and $\sigma_n=0.1$, respectively.

From Fig.~\ref{fig:coprime1} and Fig.~\ref{fig:coprime2}, we observe that OMP cannot resolve the third and fourth rays. 
OMP also misses the fifth ray when $\sigma_n=0.1$. In addition, there are quite a few side lobes that will be interpreted as targets in radar applications. Note that the iteration of OMP stops at $15$ in this simulation because the number of measurements $M=16$.
Both CG and SBL can resolve the six rays but have large side lobes.
The side lobes of CG are larger than those of SBL. 
Remarkably, BLRC suppresses spurious targets well.

Similar to SPA, the sparse signal recovery properties of SBL and BLRC using CPA can be seen from the two bottom subplots in  Fig.~\ref{fig:NR3_CPA_convergent}.
For SBL in the bottom right subplot, the six top curves corresponding to the six largest  $\tilde{\sigma}_i$ represent the six true targets.
The rest of the curves with smaller $\tilde{\sigma}_i$'s represent the false targets.
However, as the $\tilde{\sigma}_i$'s representing some of the false targets are only slightly smaller than the $\tilde{\sigma}_i$'s representing the true targets, the false targets are almost as strong as the true targets.
For BLRC in the bottom left subplot, $\hat{\gamma}$ decreases so as to promote sparsity as $k$ increases.
The implication of reducing $\hat{\gamma}^{(k)}$ is shown in Fig.~\ref{fig:GC}, which has been discussed previously for SPA and will not be repeated here.

The sparse signal recovery properties of SBL and BLRC using CPA can also be seen from the two bottom subplots in Fig.~\ref{fig:weightc}.
The first six points with $i=1,2,...,6$ with the six smallest weights represent the six true targets.
The discussions are the same as those for SPA and are omitted here.
Again, SBL has many more spurious targets than BLRC.

When $\sigma_n$ increases from $0.01$ to $0.1$, the magnitudes of false targets derived from OMP, SBL, and CG increase  (comparing Fig.~\ref{fig:coprime1} with Fig.~\ref{fig:coprime2}).
However, the spurious targets are still suppressed by BLRC. 
It is concluded that BLRC can suppress spurious targets in moderate SNR's even with a small array ($M=16$).

In summary, BLRC provides sparser solution than SBL with sparse array input. This outcome can be attributed to BLRC’s ability to generate a more accurate estimation of the noise standard deviation, $\sigma_n$, while SBL tends to produce a much smaller value for $\sigma_n$ estimation. (Further details can be found in the subsequent section.) This issue is exacerbated when sparse array input is utilized, as non-uniform random sampling further worsens the ambiguity problem. From (\ref{eq:joint_pdf}) and (\ref{eq:sbl_IR}), we know that $\sigma_n^2$ is the weight on the penalty term. When $\sigma_n^2$ converges to a reasonable non-zero value, further decrement of $\hat{\gamma}^{(k)}$ in BLRC will make $\ln(1+\frac{|c|^2}{\hat{\gamma}^2})$ in (\ref{eq:joint_pdf}) approach $||c||_0$, i.e., the sparest solution.   

\subsection{Noise Standard Deviation Estimation}
In the top right subplots in Fig.~\ref{fig:NR1_2_2} and Fig.~\ref{fig:NR3_CPA_convergent} for SPA and CPA, respectively, we show the noise standard deviations in dB ($20 \log_{10} \tilde{\sigma}_n^{(k)}$ for SBL and 
$20 \log_{10} \hat{\sigma}_n^{(k)}$ for BLRC) as functions of the iteration index $k$.
As the iteration number $k$ increases, $\tilde{\sigma}_n^{(k)}$ for SBL decreases to  very small values.
However, unlike SBL, BLRC provides a reasonable estimate of the noise standard deviation. Note that the estimated
$\hat{\sigma}_n$ is the combined effect of the original additive white Gaussian noise (AWGN) and the insufficient-sampling noise  \cite{stankovic14}. 
Thus, $\hat{\sigma}_n$ estimated by BLRC is greater than the AWGN ${\sigma}_n$. 
When ${\sigma}_n$ is large, it will be shown in the next section that the estimated $\hat{\sigma}_n \approx \sigma_n$ because the sampling noise now is negligible compared to AWGN (see the right subplot in Fig.~\ref{fig:NR1_5}).

From (\ref{eq:sbl_IR}), one can see that SBL with a very small $\tilde{\sigma}_n^{(k)}$ de-emphasizes the regularization term and therefore does not promotes sparsity at the end of the iterative process (see the top right subplots in Fig.~\ref{fig:NR1_2_2} and Fig.~\ref{fig:NR3_CPA_convergent}).
However, from (\ref{eq:joint_pdf}), one can see that BLRC promotes sparsity since the regularization term remains effective in the entire iteration process as  $\hat{\sigma}_n \ge \sigma_n$.

To understand why BLRC has a superior performance over SBL in estimating $\sigma_n$, rewrite the $\sigma_n^2$ updating strategy of BLRC in step 9 of Algorithm 3 as:
\begin{equation}
    \begin{aligned}
    \hat{\sigma}_n^2
    &=\frac{||\mathbf{y-A}\mathbf{\hat{c}} ||^2}
    {M-\text{Tr}(\mathbf{I}-\frac{2}{\hat{\gamma}^2}\hat{\mathbf{Q}} \hat{\mathbf{\Gamma}})}
    \end{aligned}
    \label{eq:noise}
\end{equation}
As shown in Fig.~\ref{fig:weightc},
$\frac{2}{\hat{\gamma}^2}\hat{\mathbf{Q}}$ in (\ref{eq:noise}) is approximately a constant diagonal loading matrix of $\mathbf{H}_{blrc}$ where
$\hat{\mathbf{\Gamma}}=\mathbf{H}_{blrc}^{-1}$.
Since there are $N-K$ very large constant diagonal terms in
$\frac{2}{\hat{\gamma}^2}\hat{\mathbf{Q}}$, $\mathbf{H}_{blrc}$ is approximately equal to $\frac{2}{\hat{\gamma}^2}\hat{\mathbf{Q}} $.
Thus, 
$\frac{2}{\hat{\gamma}^2}\hat{\mathbf{Q}} \hat{\mathbf{\Gamma}}$ in (\ref{eq:noise}) is approximately a diagonal matrix with $(N-K)$ $1$'s and $(K)$ $0$'s.
Then, the denominator in (\ref{eq:noise}) is approximately equal to $M-K$.
As the numerator in (\ref{eq:noise}) is approximately equal to $M {\sigma}_n^2$ if $\mathbf{\hat{c}}\approx \mathbf{c}$, we have $ \hat{\sigma}_n^2 \approx \frac{M}{M-K} \sigma_n^2$.

Even though the $\sigma_n^2$ updating strategy of SBL in step 6 of Algorithm 2:
\begin{equation}
    \tilde{\sigma}_n^2=\frac{|| \mathbf{y}-\mathbf{A}\tilde{\mathbf{c}}||^2}{M-\text{Tr}(\mathbf{I}-\tilde{\mathbf{\Gamma}}\tilde{\mathbf{\Sigma}})}
\end{equation}
is of the same form as  (\ref{eq:noise}),
$\tilde{\mathbf{\Sigma}}$ is nearly singular and its diagonal terms are far from constant.
Thus, SBL does not estimate $\sigma_n^2$ accurately.
In fact, as $\tilde{{\sigma}}_n^{(k)}$ becomes smaller,  the penalty term for non-sparsity in (\ref{eq:sbl_IR}) becomes smaller as well. Then, as the iteration proceeds, the data fitting term $||\mathbf{y-Ac}||^2$  in (\ref{eq:sbl_IR}) becomes more and more important and the estimated $\tilde{{\sigma}}_n^{(k)}$ becomes disproportionately small.

\subsection{Sensitivity with respect to $\sigma_n$, $K$ and $M$}

\begin{figure}[!t]
\centering
  \includegraphics[width=3.4in]{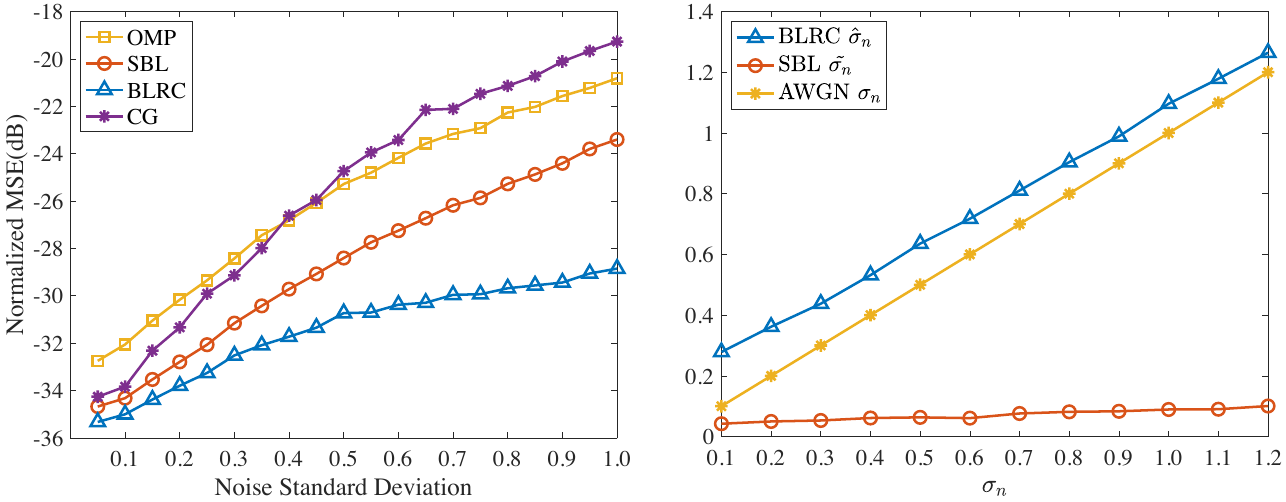}
  \caption{(Left) The normalized MSE as noise standard deviation increases; (Right) The estimated $\sigma_n$ by BLRC and SBL }
  \label{fig:NR1_5}
\end{figure}

\begin{figure}[!t]
\centering
  \includegraphics[width=3.4in]{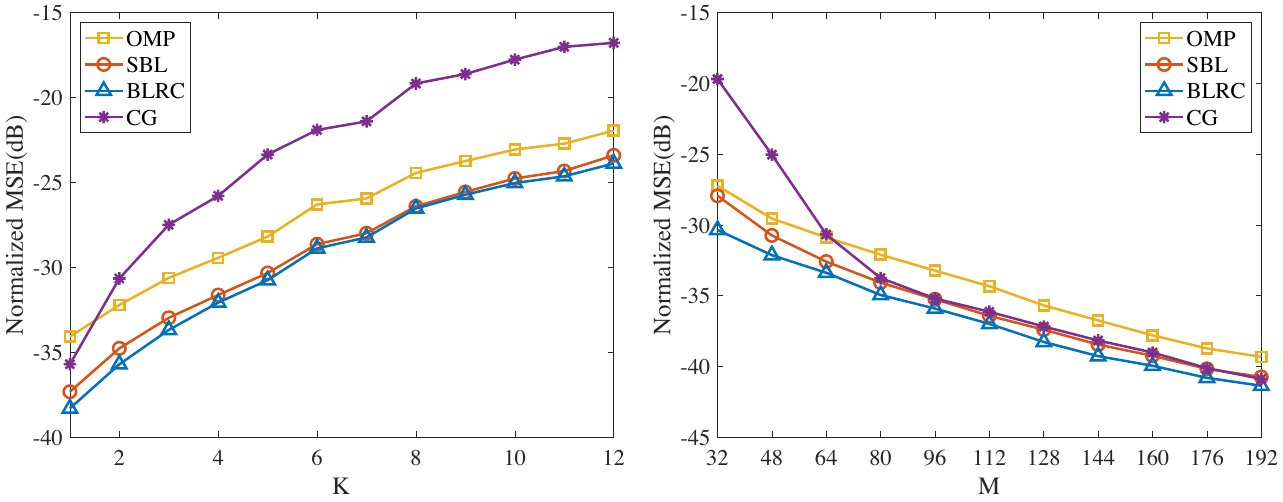}
  \caption{(Left) The normalized MSE as the number of targets $K$ changes; (Right) The normalized MSE as the number of measurements M changes}
  \label{fig:MSE_KM}
\end{figure}

Sensitivities of the four spectrum reconstruction results (derived by OMP, SBL, CG, and BLRC) with respect to the noise standard deviation $\sigma_n$,  number of targets $K$, and number of samples $M$ are shown
in Fig. \ref{fig:NR1_5} and Fig. \ref{fig:MSE_KM}.
In these figures, the normalized mean squared error (MSE) is defined as
\begin{equation}
    \text{MSE (dB)} =10\log_{10} \Big[ \frac{1}{N}|| \frac{{\mathbf{c}}}{\max_i\{|{c}_i|\}}-\frac{\bar{\mathbf{c}}}{\max_i\{|\bar{c}_i|\}}  ||_2^2 \Big]
    \label{eq:mse}
\end{equation}
where $N=256$ and $\mathbf{c}$ is the true complex spectrum of the receive signal and
$\bar{\mathbf{c}}$ represents the recovered spectrum using OMP, SBL, CG, or BLRC. 
Note that MSE may not reflect the spectrum reconstruction well when 
the estimated ray frequencies are slightly off from the corresponding true ray frequencies. 
In addition, MSE does not show sparsity characteristics.
Thus, this simulation is conducted only for SPA with large $M$ where sparse signal recovery can be done relatively easily. 

\begin{table}
\caption{Parameter Settings}
\label{table}
\tablefont
\begin{tabular*}{20pc}{@{}c@{}}
%\hline
\centerline{\includegraphics[width=20pc]{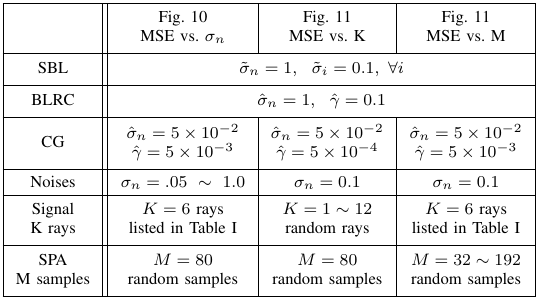}}\\
%\hline
\end{tabular*}
\label{tab1}
\end{table}

The MSE at each point plotted in Fig. \ref{fig:NR1_5} and Fig. \ref{fig:MSE_KM}
is the average of 100 realizations of random noises, random rays and random samples, respectively.
The parameters for the three Bayesian approaches (SBL, CG, and BLRC) are outlined in Table~\ref{tab:y}.
The number of samples,  ray parameters, and  noise standard deviations are also shown in Table~\ref{tab:y}.

The average MSE's of 100 realizations obtained by OMP, SBL, CG, and BLRC as noise standard deviation increases from 0.05 to 1 are shown in the left subplot of Fig.~\ref{fig:NR1_5}. BLRC shows the best performance, which is followed by SBL.
CG performs better than OMP when noise is small. 
But CG performs worse than OMP with large noises,
which is due to the fact that the parameters of CG are not adjusted as noise increases.

The right subplot in Fig.~\ref{fig:NR1_5} shows the average estimated noise standard deviations of 100 realizations obtained by BLRC and SBL for several different noise levels. 
It can be seen that the SBL estimates are inaccurate.
On the other hand, BLRC estimates are larger than, but very close to, the true noise values.
The higher the noise level is, the more accurate the BLRC estimate is.
This is because, as mentioned before, there are mainly two kinds of noise in sparse signal recovery problems \cite{stankovic14}: additive white Gaussian noise (AWGN) and random sampling noise. 
What BLRC estimate is the combined effect of these two kinds of noise.
When AWGN is large, AWGN becomes dominant and the random sampling noise becomes negligible.

The average MSE's of 100 realizations obtained by OMP, SBL, CG and BLRC as the number of rays, $K$, increases from 1 to 12 are shown in the left subplot in Fig.~\ref{fig:MSE_KM}. 
Table~\ref{tab:y} shows the parameters used for the simulations.
Similarly, BLRC shows the best performance followed by SBL.
CG performs better than OMP when $K=1$. 
But CG performs worse than OMP for all other $K$'s,
which is due to the fact that optimum CG parameters used for simulation need to be found by trial and error in order to get good performances.
Here, the parameters are chosen for $K=1$ and not adjusted further as $K$ increases.

The average MSE's of 100 realizations obtained by OMP, SBL, CG, and BLRC as the number of SPA elements, $M$, increases from 32 to 192 are shown in the right subplot in Fig.~\ref{fig:MSE_KM}. 
Table~\ref{tab:y} shows the parameters used for simulations.
BLRC has the best performance and SBL is the second-best.
CG performs better than OMP when $M$ is large. 
But CG performs worse than OMP for small $M$'s.
Again, the parameters of CG are not adjusted as $M$ changes. 

\subsection{Resolution}

Using SPA with $M=80$, OMP, CG, SBL, and BLRC have very similar resolution performances. Their estimated results are all accurate because the aperture of SPA with $M=80$ is large enough and sufficient measurements are obtained. Here, the CPA with $M=16$ is used to test the resolution of the proposed BLRC approach, as high resolution cannot be achieved easily with such a small aperture. 
It will be shown that BLRC outperforms OMP, CG, and SBL.

\begin{figure}[!t]
\centering
  \includegraphics[width=3.4in]{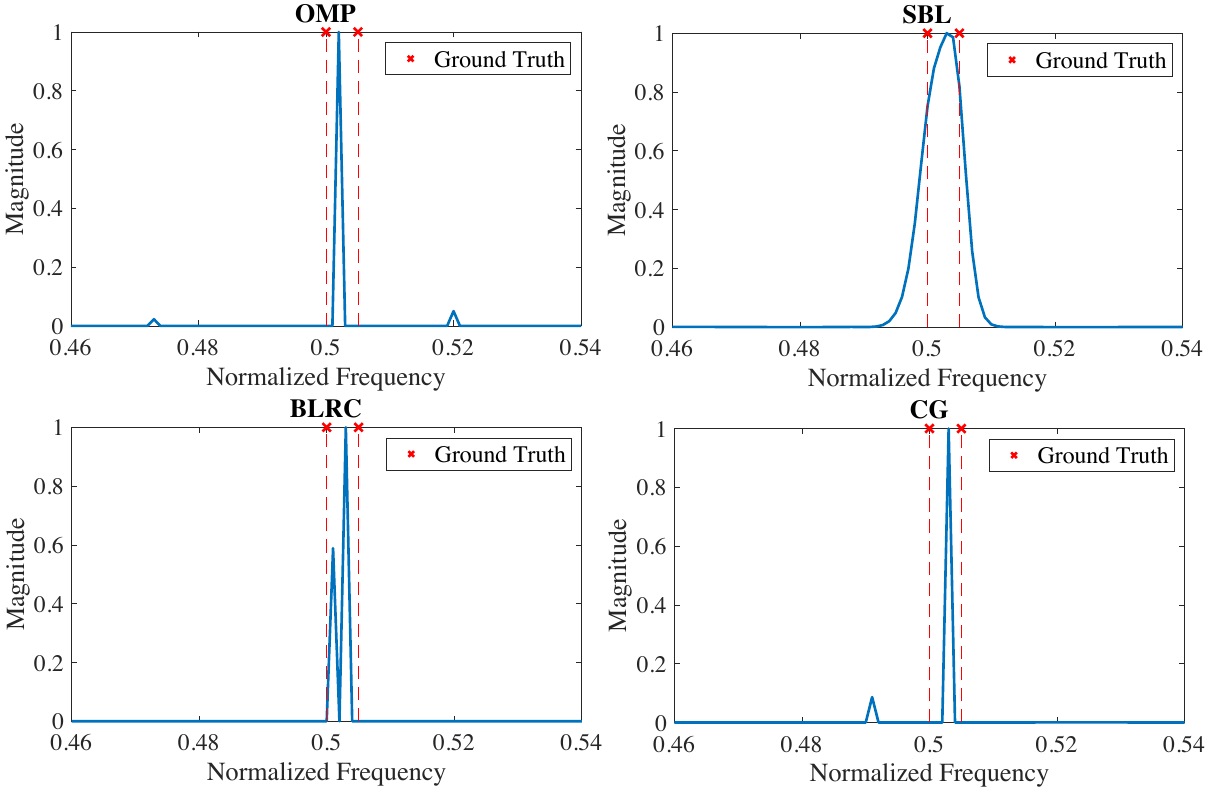}
  \caption{Resolution of Two Rays of Same Magnitude}
  \label{fig:resolution1}
\end{figure}

In the first example, consider two rays with equal amplitudes. Their normalized frequencies are $\frac{500}{N}$ and $\frac{505}{N}$ where $N=1000$. 
Fig.~\ref{fig:resolution1} shows that OMP, SBL and CG all fail to resolve the two rays while BLRC can  distinguish them. Note that SBL stops at iteration 6 in this example because the matrix to be inverted becomes singular at iteration 7 (see numerical stability in Section VI.C).

\begin{figure}[!t]
\centering
  \includegraphics[width=3.3in]{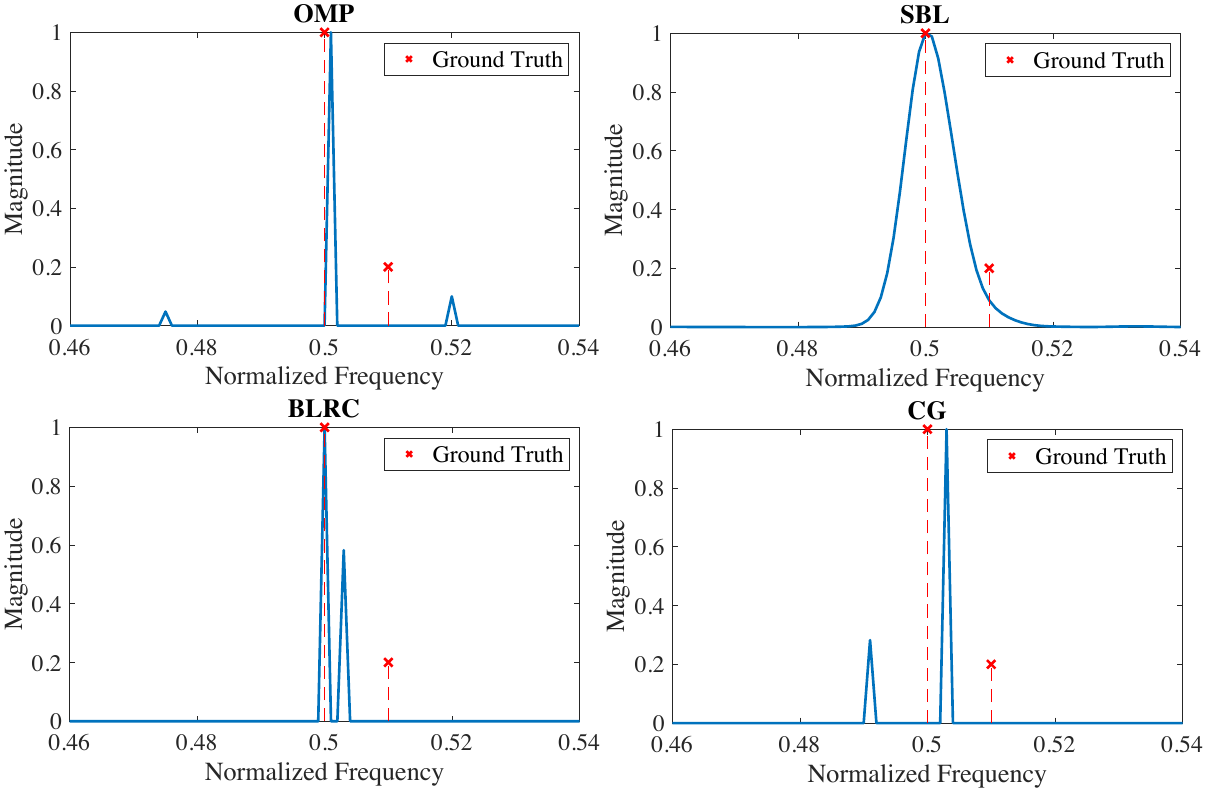}
  \caption{Resolution of Two Rays with Different Magnitudes}
  \label{fig:resolution2}
\end{figure}

In the second example, consider two rays with different magnitudes: 1 and 0.2. Let their normalized frequencies be $\frac{500}{N}$ and $\frac{510}{N}$, respectively, where $N=1000$.  Again, Fig.~\ref{fig:resolution2} shows that OMP, SBL and CG all fail to resolve the two rays while BLRC can distinguish them. Similarly, SBL stops at iteration 6 due to the near singular issue. 

Although the two rays can be distinguished by BLRC in both examples, the estimated frequencies and magnitudes are slightly off because the targets are so close to each other. Even worse when one is weaker than the other, the weak target estimation is biased toward the strong target. Fortunately, for current autonomous driving applications, these results are satisfactory because the ability to separate targets is more crucial (e.g., early detection of crossing pedestrians \cite{palffy2019occlusion}).

Note that in both Fig.~\ref{fig:resolution1} and Fig.~\ref{fig:resolution2}, SBL has spurious targets while BLRC does not. This is not shown in these two zoom-in figures.
The better performances of BLRC over SBL on resolution and sparsity are due to the fact that BLRC provides a reasonable estimation of the noise standard deviation $\sigma_n^2$, but SBL does not (as discussed in Sections VI.F and VI.G).

\section{Automotive Radar Applications}

In this section, simulated automotive radar signals are processed to demonstrate the performance of the proposed BLRC method. 

\subsection{Image Radar Signal Processing}

In order to locate targets in 2D, we need to receive the back scattered electromagnetic signal in two distinct dimensions \cite{automotiveDSP,MS92,PS13}. Consider the following frequency-modulated continuous-wave (FMCW) transmitted pulse:
\begin{equation}
    s(t) = e^{j2\pi (f_c+\frac{\alpha t}{2})t},~0\leq t\leq T_c
    \label{eq:chirp}
\end{equation}
where $f_c$ is the carrier frequency, $\alpha$ is the chirp slope, and $T_c$ is the duration of one chirp.

\begin{figure}[!t]
\centering
  \includegraphics[width=3.3in]{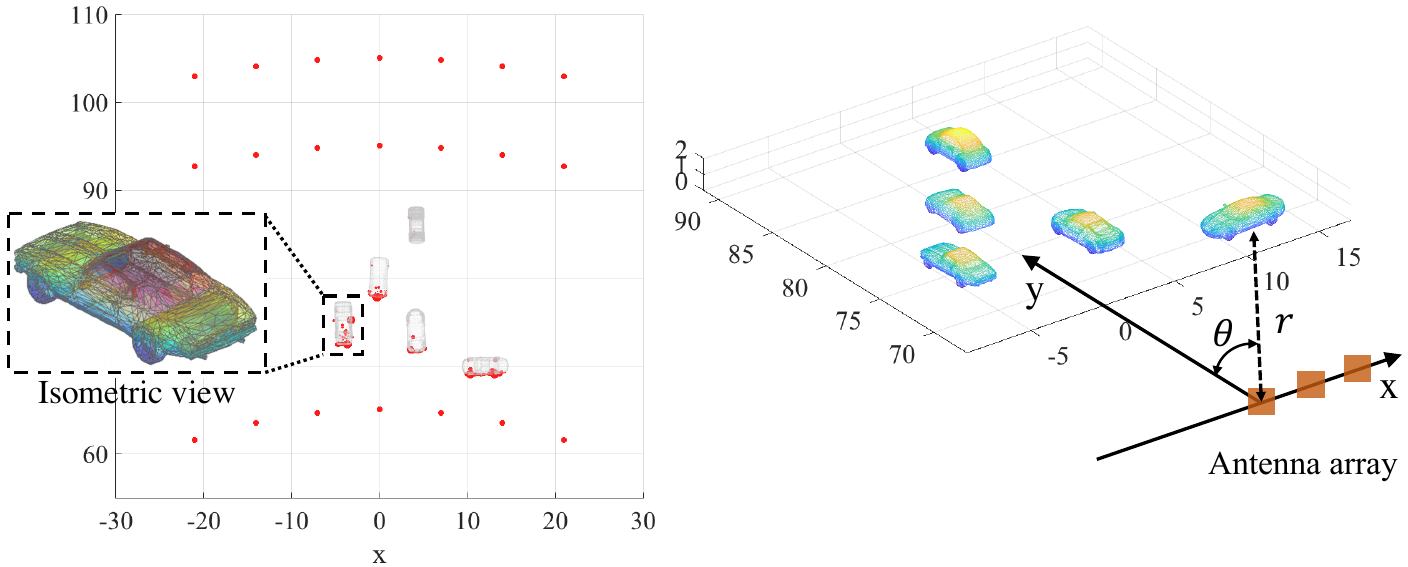}
  \caption{Simulation scenario for range-azimuth angle estimation (top view and side view)}
  \label{fig:rtheta}
\end{figure}

Consider Fig.~\ref{fig:rtheta} as an example where the first antenna array element is set as the origin.
Since the targets are in the far zone, plane wave approximation is valid. 
Define $P(r_l,\theta_l)$ as the position of $l^{th}$ target in the polar coordinates, then the round-trip propagation time between 
the $l^{th}$ target and the $i^{th}$ array element is given by
\begin{equation}
    t_l = \frac{2r_l-id\sin\theta_l}{v}
\end{equation}
where $d$ is the antenna spacing and $v$ is the speed of light.

To estimate the range $\{r_l \}$ and angle $\{\theta_l \}$, the $t_l$-delayed received signal at the $i^{th}$ antenna element is mixed with $s^*(t)$ by the mixer, filtered by low-pass filter (LPF), and sampled at the sampling rate $f_s$ by analog-to-digital converter (ADC).
Then, the $n^{th}$ ADC samples at the $i^{th}$ antenna element is: 
\begin{equation}
\begin{aligned}
    \check y(n,i)&\approx\sum_{l=1}^K \check{a}_l \exp \Big\{j2\pi \big[ \frac{2r_l}{v}\alpha n\Delta t -\frac{d\sin\theta_l}{\lambda}i +\frac{2r_l}{\lambda}\big]      \Big\}\\
    &+\check\epsilon(n,i), ~~~~~~~~~~~~~~~n=1,2,...,N_s
    \label{eq:adc}
\end{aligned}
\end{equation}
where $K$ is the number of targets, $\Delta t =1/f_s$ is the sampling interval, $N_s$ is the number of ADC samples, $\check\epsilon$ is the additive noise and $\Check{a}_l$ is the $l^{th}$ target strength.
Note that the received target strength $|\Check{a}_l|$ is proportional to the antenna gain, RCS of the target and $1/r_l$. 

To obtain the range estimation, performs FFT on the $N_s$ ADC samples in (\ref{eq:adc}):
\begin{equation}
\begin{aligned}
    {y}(p,i)=\mathcal{F}_n\{ \check y(n,i) \}
    =\sum_{n=0}^{N_s-1} \check y(n,i)e^{-j2\pi \frac{p}{N_s}n}
       \label{eq:angle}
\end{aligned}
\end{equation} 
Representing (\ref{eq:angle}) in the same form as (\ref{eq:example_1}) with full samples, we have
\begin{equation}
\begin{aligned}
     {y}(p,i)=\sum_{l=1}^{K}  a_l (p)\exp\{j (2 \pi f_l \frac{i}{N}  + \phi_l)\}+\epsilon(p,i)
    \label{eq:angle_1}
\end{aligned}
\end{equation}
where $a_l(p)=\mathcal{F}_n \Big\{ \check{a}_l e^{j2\pi \big[ \frac{2r_l}{v}\alpha n\Delta t  \big]  }  \Big\}$,
$f_l=-\frac{L\sin\theta_l}{\lambda}$, 
$\phi_l=\frac{4\pi r_l}{\lambda}$ and
$\epsilon(p,i)=\mathcal{F}_n\{ \check \epsilon(n,i) \}$. Note that the $d$ in (\ref{eq:adc}) is the minimum element spacing in antenna array and $d=\frac{L}{N}$ where $L$ is the full antenna aperture if none of the $N$ array elements is missing. Therefore, the SPA and CPA used in previous sections can be obtained by selecting array elements based on the corresponding pattern as shown in Fig.~\ref{fig:dft_omp3}. 

Note that, when $p=\frac{2r_l}{v}\alpha N_s\Delta t$, $|a_l(p)|$ gets its maximum value: $N_s |\check{a}_l|$. 
Usually, if the maximum of $|a_l(p)|$ occurs at $p=p^*$, a local maximum of $|{y}(p,i)|$ also occurs at $p=p^*$.
Thus, the indexes of local maximums of $|{y}(p,i)|$ offer the range information 
$\{\frac{2r_l}{v}\alpha N_s\Delta t\}$. 

To find the angle information, we can apply sparse spectrum reconstruction approaches (with respect to $i$) to $y(p^*,i)$ in (\ref{eq:angle_1}). For automotive radar applications, the number of samples $M$ which is usually much less than $N$.
Since the number of dominant targets for a given range is small, there will be no problem to find the spatial frequency $f_l$ corresponding to $p^*$ in (\ref{eq:angle_1}). 
From $f_l$ and $p^*$, we can then find the angle and range of target, respectively. Note that, without loss of generality, the Doppler processing is omitted for simplicity.

\subsection{Automotive Radar Signal Simulator}
A physical optics (PO) based electromagnetic simulator inspired by \cite{PO} is developed to generate typical FMCW automotive radar signals.
In our simulator, targets are composed of many small triangular facets. To generate the received radar signal, the simulator at first computes the induced currents on each facet of all targets illuminated by the incident radar signal based on the principle of PO.  
Then, the simulator sums up, at the radar receive antenna, all back-scattered electromagnetic fields radiated from these induced currents on all facets of all targets.
Using this simulator, scattered electromagnetic fields and radar cross section (RCS) of each target in realistic target scenes can be generated.
Compared to the computationally intensive and resource consuming full-wave simulation, the PO based simulator is able to highlight dominant wave features and provide computational efficiency without compromising much on numerical accuracy if the high frequency approximation for electromagnetic wave propagation and scattering is valid.

\subsection{Numerical Example}

An automotive radar with a 16-element coprime array (as marked in the top subplot of Fig.~\ref{fig:dft_omp3}) is considered. A scene with five cars and twenty-one fixed point targets (corner reflectors) 
is shown in the left subfigure of Fig.~\ref{fig:rtheta}. The first element of the coprime array is located at the origin as shown in the right subfigure of Fig.~\ref{fig:rtheta}. Specific parameters for all these targets and the radar are shown in Table~\ref{tab:radar}. 

\begin{table}[!t]
\renewcommand{\arraystretch}{1.3}
\centering
\caption{\textsc{Radar Setting}}
\label{tab:radar}
\begin{tabular}{|c||c|}
%\midrule
\hline
Parameter  & Setting  \\
\hline
Carrier Frequency $f_c$ & 79GHz   \\
\hline
Wavelength $\lambda$ & 3.797 mm  \\
\hline
Chirp Slope $\alpha$ & 10MHz/$\mu$sec  \\
\hline
Sampling Frequency $f_s$ & 20MHz\\
\hline
ADC Samples $N_s$ & 1024\\
\hline
Antenna Number $M$ & 16\\
\hline
Antenna Element Spacing $d$ & $\lambda/2$\\
\hline
Target Surface Resistivity & 0.1 \\
\hline
Car1 Position & [12,70,0] ~in~meters\\
\hline
Car2 Position &[-4.1,75,0] ~in~meters\\
\hline
Car3 Position & [-0.1,80.1,0]~in~meters\\
\hline
Car4 Position & [4.21,86,0] ~in~meters\\
\hline
Car5 Position & [4.15,74,0] ~in~meters\\
\hline
Fixed Target Ranges & 65,95,105 ~in~meters\\
\hline 
Fixed Target Angles & -21,-14,-7,0,7,14,21 ~in~degrees\\
\hline
\end{tabular}
\end{table}

Then FMCW signal in (\ref{eq:chirp}) with a chirp rate 10MHz/$\mu$sec is transmitted. 
In our simulator, all targets are composed of many small facets. 
For example, the automobile in isometric view in Fig.~\ref{fig:rtheta} is composed of 7226 facets.
Then we use our proposed automotive radar signal simulator to generate the received signals at the 16 coprime array elements. The additive white Gaussian noise (AWGN) $\check{\epsilon}$ in (\ref{eq:adc}) is generated with two different standard deviations, $\sigma=$ 0.03 and $\sigma=$ 0.3.

\begin{figure*}[!t]
\centering
\includegraphics[width=6.8in]{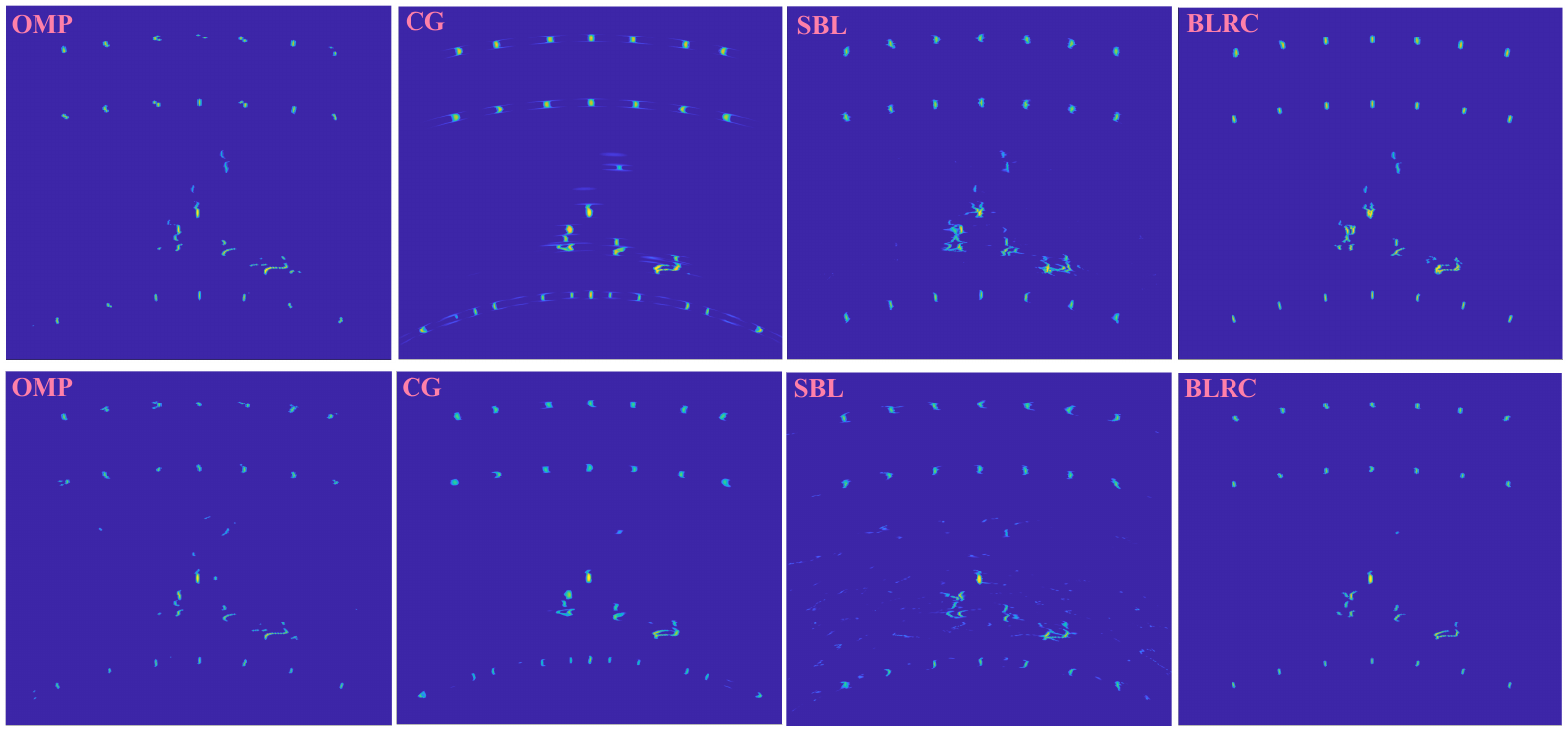}
%\hfil
% where an .eps filename suffix will be assumed under latex, 
% and a .pdf suffix will be assumed for pdflatex; or what has been declared
% via \DeclareGraphicsExtensions.
\caption{Radar Imaging Performances for OMP, CG, SBL, and BLRC based on Realistic Target Scene Simulation (Image Dynamic Range: 70dB). Upper Row: $\sigma= 0.03$. Lower Row: $\sigma= 0.3$.}
\label{fig:2Dresult}
\end{figure*}

After mixer, LPF, and ADC, the signal received by the $i^{th}$ array element becomes the ADC sample $\check y(n,i)$ in (\ref{eq:adc}).
Perform FFT on $\check y(n,i)$ with respect to $n$ to obtain 
$y(p,i)=\mathcal{F}_n\{ \check y(n,i) \}$ as shown in (\ref{eq:angle}).  
Then, perform OMP, CG, SBL, and BLRC on $y(p,i)$ with respect to $i$ to obtain
the 2D radar spectrum $\bar{y}(p,f)$ for each of these four approaches.
Using $r=\frac{pv}{2 \alpha N_s \Delta t}$ to convert $p$ to the range variable $r$ and using $\theta = \arcsin{\frac{-\lambda f}{L}}$ to map $f$ to the angle $\theta$, the 2D radar image plot $|\bar{y}(r,\theta)|$ recovered by the four approaches mentioned in Section VI with two different AWGN strengths are shown in Fig.~\ref{fig:2Dresult}. 

As shown in Fig.~\ref{fig:2Dresult}, with pruning, SBL results still exhibit spurious targets. Moreover, the number of spurious targets increases with noise strength. CG results also show spurious targets at range equal to $65~m$, even though the CG parameters have been optimized by trial and error. Note that optimizing parameters by trial and error is not possible in practice, since the targets are supposed to be unknown.
For OMP, we set the iteration number to be $8$, which is slightly larger than the maximum target number at each range. This is not possible in practical applications either because the targets are unknown. 
It is remarkable that, BLRC robustly delivers the least spurious solution and provides optimal resolution of the target images for both AWGN strengths.
This is consistent with our previous analyses.

\section{Conclusion}

In this paper, we propose a Bayesian linear regression algorithm, BLRC, which uses the non-conjugate Cauchy prior. Then we focus on the comparisons among three Bayesian linear regression approaches (i.e., BLRC, CG, and SBL) when sparse array measurement is used as input.

Firstly, BLRC can be considered as a significant improvement over the CG approach 
because BLRC provides a systematic updating scheme for the hyper-parameters which is absent from the CG approach.
This makes BLRC more feasible than CG as CG is sensitive to the selection of the hyper-parameters. Furthermore, the systematic updating scheme empowers BLRC the capability to reduce the number of local minimums as shown in Section IV. This greatly reduces the chances of BLRC being trapped in a local minimum at the early stage of the iteration process. As is shown in Section VI and Section VII, BLRC outperforms CG in various scenarios even when CG has the best choice of its hyper-parameters. This makes BLRC more practical than CG, especially when it is used in the highly dynamic automotive scenarios.

Secondly, BLRC can also be considered as a significant improvement over the well-known SBL approach. 
Both BLRC and SBL are Bayesian linear regression approaches and have the same iterative updating steps for the solution and hyper-parameters. The formulas for updating the solutions in BLRC and SBL are also similar to each other. However, there are key differences between BLRC and SBL. Only two hyper-parameters need to be handled in BLRC due to the use of Cauchy prior, while the number of hyper-parameters to be handled is large for SBL. Thus, compared with SBL, BLRC has a more compact latent space. 

Thirdy, comprehensive numerical analyses are conducted to demonstrate the superior performances of BLRC when sparse array measurement is used as input. Special attention has been paid to the comparisons between BLRC and the well-known SBL approach. It is shown that BLRC is more numerically robust than SBL and provides a more accurate estimate of the noise variance. Since BLRC provides a more accurate estimation of the noise variance, it tends to yield the sparsest solution, whereas SBL is susceptible to generating spurious targets. The issue of spurious targets becomes more pronounced in scenarios involving sparse array or low-SNR input, as both factors contribute to exacerbating the ambiguity problem. From the resolution point of view, the sparest solution tends to separate targets even though they are very close to each other. Thus, BLRC provides a higher resolution result. Based on the IR-$l_2$ interpretations, it is remarkable to see that all advantages of BLRC over SBL are originated from the simple fact that BLRC uses the long-tailed Cauchy prior which leads to a more compact latent space.

Finally, the application of BLRC to sparse MIMO radar array signal processing is presented. Compared with the reconstructed radar images of other sparse signal recovery algorithms, the performance of BLRC successfully demonstrates its efficiency in producing high-resolution radar images with the least false targets, which is critical to the development of advanced driver-assistance systems (ADAS) and autonomous driving (AD) applications.

\appendices
\section{}
The first expectation in (\ref{eq:expectation}) is further simplified below:
\begin{equation}
    \begin{aligned}
    E_{\mathbf{c|y}}[||\mathbf{y-Ac} ||^2]&=
    |\mathbf{y}|^2
    - \mathbf{\hat{c}}^T \mathbf{A}^T \mathbf{y}
    -\mathbf{y}^T \mathbf{A} \mathbf{\hat{c}}\\ 
    &+E_{\mathbf{c|y}}[\mathbf{c}^T \mathbf{A}^T\mathbf{A}\mathbf{c}]
    \label{eq:exp_1}
    \end{aligned}
\end{equation}
As $\mathbf{c}^T \mathbf{A}^T\mathbf{A}\mathbf{c}$ is a scalar, 
\[
\mathbf{c}^T \mathbf{A}^T\mathbf{A}\mathbf{c}
=\text{Tr}(\mathbf{c}^T \mathbf{A}^T\mathbf{A}\mathbf{c})=\text{Tr}(\mathbf{A}\mathbf{c}\mathbf{c}^T \mathbf{A}^T)\\
\]
Then,
\begin{equation}
\begin{aligned}
E_{\mathbf{c|y}}[\mathbf{c}^T \mathbf{A}^T\mathbf{A}\mathbf{c}] &=
\text{Tr}(\mathbf{A} E_{\mathbf{c|y}}[\mathbf{c}\mathbf{c}^T] \mathbf{A}^T)
  \\
  &=\text{Tr}(\mathbf{A} \mathbf{\hat{\Gamma}} \mathbf{A}^T) 
  +\text{Tr}(\mathbf{A} \mathbf{\hat{c}}\mathbf{\hat{c}}^T\mathbf{A}^T)\\   
    &=\text{Tr}(\mathbf{A}^T \mathbf{A} \mathbf{\hat{\Gamma}} )
  +\mathbf{\hat{c}}^T \mathbf{A}^T \mathbf{A} \mathbf{\hat{c}}.
\end{aligned}
\end{equation}
Substituting the above equation into (\ref{eq:exp_1}), we have
\begin{equation}
    \begin{aligned}
    E_{\mathbf{c|y}}[||\mathbf{y-Ac} ||^2]
    =||\mathbf{y-A\hat{\mathbf{c}}}  ||^2+ \text{Tr}(\mathbf{A}^T\mathbf{A}\hat{\mathbf{\Gamma}})
    %\label{eq:exp_2}
    \end{aligned}
\end{equation}

\section{}
Set $a\rightarrow 0$ and $b\rightarrow 0$, then we extend the lower bound in (\ref{eq:convexbound}) of a scalar $c$ to a lower bound of a vector $\mathbf{c}$, and use the notation $\tau_i$ instead of $\xi_i$:
\begin{equation}
\begin{aligned}
     p(\mathbf{c}|\pmb{\tau})=\prod_{i=1}^N p(c_i)
     \ge \mathcal{N}(\mathbf{0},\mathbf{\Sigma}^{-1})
\end{aligned}
\label{eq:pctau}
\end{equation}
where $\mathbf{\Sigma}$ has been defined as $\text{diag}(\pmb{\tau})$ in (\ref{eq:sbl_prior}).
In (\ref{eq:pctau}), the dependence on $\pmb{\tau}$ is explicitly shown. 
Recall the optimization problem for hyper-parameters in (\ref{eq:hier1}):
\begin{equation}
    \begin{aligned}
        \{\tilde{\pmb{\tau}},\tilde{\sigma}_n\} &=\arg\max_{\pmb{\tau},\sigma_n}
         \ln \{ p(\mathbf{y}|\pmb{\tau},\sigma_n) \}\\
         &=\arg\max_{\pmb{\tau},\sigma_n}
         \ln \{ \int p(\mathbf{y}|\mathbf{c},\sigma_n)p(\mathbf{c}|\pmb{\tau})d\mathbf{c} \}
         \label{eq:opt_sbl}
\end{aligned}
\end{equation}
Substituting (\ref{eq:likelihood}) and (\ref{eq:pctau}) into the integral in (\ref{eq:opt_sbl}), the lower bound of the evidence distribution $p(\mathbf{y}|\pmb{\tau},\sigma_n)$ is derived as follows:
\begin{equation}
    \begin{aligned}
        p(\mathbf{y}|\pmb{\tau},\sigma_n) 
        \ge &\int \mathcal{N}(\mathbf{Ac},\sigma_n^2\mathbf{I})\mathcal{N}(\mathbf{0},\mathbf{\Sigma}^{-1})d\mathbf{c}
        =\mathcal{N}(\mathbf{0},\mathbf{\Sigma}_y^{-1})
    \end{aligned}
\end{equation}
with
\begin{equation}
 \mathbf{\Sigma}_y^{-1}=\sigma_n^2\mathbf{I}+\mathbf{A}\mathbf{\Sigma}^{-1}\mathbf{A}^T 
\end{equation}

According the equation (36) in \cite{sblrvm}, the following is the negative logarithm of the lower bound of $p(\mathbf{y}| \pmb{\tau}, \sigma_n)$:
\begin{equation}
    \begin{aligned} 
    L_{sbl}=\ln |\mathbf{\Sigma}_y^{-1}|+\mathbf{y}^T\mathbf{\Sigma}_y\mathbf{y}+C
        \label{eq:L_sbl0}
    \end{aligned}
\end{equation}
where $C$ is a constant.
Thus, optimization problem in (\ref{eq:opt_sbl}) is equivalent to minimizing the cost function $L_{sbl}$ in (\ref{eq:L_sbl0}).

As shown in Appendix C, 
\begin{equation}
    \begin{aligned}
        \mathbf{y}^T\mathbf{\Sigma}\mathbf{y}
        =\min_{\mathbf{c}}~ \Big(\frac{1}{\sigma_n^2}|| \mathbf{y}-\mathbf{Ac}||^2+\mathbf{c}^T\mathbf{\Sigma}\mathbf{c}  \Big)
        \label{eq:Appendix_B}
    \end{aligned}
\end{equation}
Thus, minimizing the cost function $L_{sbl}$ with respect to $\pmb{\tau}$ and $\sigma_n$ can be rewritten in $\mathbf{c}$-space as \cite{Wipf11}:
\begin{equation}
    \min_{\mathbf{c}} J_{sbl}(\mathbf{c}),~~J_{sbl}(\mathbf{c})=|| \mathbf{y}-\mathbf{Ac}||^2+h_{sbl}(\mathbf{c})
    \label{eq:L_sbl_app}
\end{equation}
where the minimization with respect to $\pmb{\tau}$ and $\sigma_n$ is done in deriving the regularization $h_{sbl}(\mathbf{c})$
to encourage a sparse solution:
\begin{equation}
    h_{sbl}(\mathbf{c})=\min_{\pmb{\tau},\sigma_n}~\sigma_n^2 \Big(\mathbf{c}^T\mathbf{\Sigma}\mathbf{c}+ \ln |\sigma_n^2\mathbf{I}+\mathbf{A}\mathbf{\Sigma}^{-1}\mathbf{A}^T| \Big)
    \label{eq:h_sbl_app}
\end{equation}

\section{}
To find the solution to
\begin{equation}
    \begin{aligned}
        \min_{\mathbf{c}}~ \Big(\frac{1}{\sigma_n^2}|| \mathbf{y}-\mathbf{Ac}||^2+\mathbf{c}^T\mathbf{\Sigma}\mathbf{c}  \Big),
        \label{eq:app_b}
    \end{aligned}
\end{equation}
set the derivative with respect to $\mathbf{c}$ to be 0:
\begin{equation}
    \frac{\partial}{\partial \mathbf{c}}\Big[\frac{1}{\sigma_n^2}|| \mathbf{y}-\mathbf{Ac}||^2+\mathbf{c}^T\mathbf{\Sigma}\mathbf{c}  \Big]_{\mathbf{c} =\mathbf{c}_o}=0
\end{equation}
so we can get the optimal $\mathbf{c}_o$:
\begin{equation}
    \mathbf{c}_o=(\sigma_n^2\mathbf{\Sigma}+\mathbf{A}^T\mathbf{A})^{-1}\mathbf{A}^T\mathbf{y}
    \label{eq:minc}
\end{equation}
Substitute $\mathbf{c}_o$ back to the original objective function in (\ref{eq:app_b}), we have
\begin{equation}
    \begin{aligned}
       &\frac{1}{\sigma_n^2}|| \mathbf{y}-\mathbf{Ac}_o||^2+\mathbf{c}_o^T\mathbf{\Sigma}\mathbf{c}_o\\
       =&\frac{1}{\sigma_n^2}\mathbf{y}^T\big[\mathbf{I}-\mathbf{A}(\sigma_n^2\mathbf{\Sigma}+\mathbf{A}^T\mathbf{A})^{-1}\mathbf{A}^T \big]\mathbf{y}.
       \label{eq:98}
    \end{aligned}
\end{equation}
Applying the Woodbury's identity
\begin{equation}
    \mathbf{E}^{-1}-\mathbf{E}^{-1}\mathbf{B}(\mathbf{D}+\mathbf{CE}^{-1}\mathbf{B})^{-1}\mathbf{CE}^{-1}
    =(\mathbf{E}+\mathbf{BD}^{-1}\mathbf{C})^{-1},
\end{equation}
to simplify $\big[\mathbf{I}-\mathbf{A}(\sigma_n^2\mathbf{\Sigma}+\mathbf{A}^T\mathbf{A})^{-1}\mathbf{A}^T \big]$ in
(\ref{eq:98}), 
we get
\begin{equation}
    \begin{aligned}
       &\frac{1}{\sigma_n^2}\mathbf{y}^T\big[\mathbf{I}-\mathbf{A}(\sigma_n^2\mathbf{\Sigma}+\mathbf{A}^T\mathbf{A})^{-1}\mathbf{A}^T \big]\mathbf{y}\\
       =&\mathbf{y}^T(\sigma_n^2\mathbf{I}+\mathbf{A}\mathbf{\Sigma}^{-1}\mathbf{A}^T)^{-1}\mathbf{y}
    \end{aligned}
\end{equation}
Thus, we have
\begin{equation}
    \begin{aligned}
        \mathbf{y}^T(\sigma_n^2\mathbf{I}+\mathbf{A}\mathbf{\Sigma}^{-1}\mathbf{A}^T)^{-1}\mathbf{y}
        =\min_{\mathbf{c}}~ \Big(\frac{1}{\sigma_n^2}|| \mathbf{y}-\mathbf{Ac}||^2+\mathbf{c}^T\mathbf{\Sigma}\mathbf{c}  \Big).
    \end{aligned}
\end{equation}

\section*{Acknowledgment}
The authors would like to thank Prof. Ivan Selesnick (New York University) and Satish Ravindran (NXP Semiconductors) for their helpful suggestions and discussions on this work. Additionally, we would like to extend our gratitude to the editor and anonymous reviewers for their valuable comments which have been instrumental in enhancing the quality and clarity of this work. We appreciate their time and effort in helping us refine this work.

%\bibsection*{REFERENCES}

\ifCLASSOPTIONcaptionsoff
  \newpage
\fi

\bibliographystyle{IEEEtaes.bst}
%\bibliography{bib-list}
\bibliography{IEEEabrv,reference}

\end{document}